\documentclass[%
preprint,
11pt,
 amsmath,amssymb,
aps,
]{revtex4-2}

\usepackage{graphicx}
\usepackage{epstopdf}
\usepackage{dcolumn}
\usepackage{bm}
\usepackage{xcolor}
\usepackage{upgreek}
\usepackage{multirow}
\usepackage{tabularx}
\usepackage[abs]{overpic}
\usepackage{placeins}
\usepackage{tikz}
\usepackage{subfig}
\usepackage{comment}

\begin{document}


\title{Asymptotic closure model for inertial particle transport in turbulent boundary layers}

\author{Y. Zhang}
\affiliation{Department of Civil and Environmental Engineering, Duke University, Durham, North Carolina 27708, USA}

\author{G. Wang}%
\affiliation{Life Science Technologies Department, Interuniversity Microelectronics Centre(IMEC), Leuven 3000, Belgium}

\author{A.D. Bragg}
\email{andrew.bragg@duke.edu}
\affiliation{Department of Civil and Environmental Engineering, Duke University, Durham, North Carolina 27708, USA}

\date{\today}

\begin{abstract}

Transport equations for heavy inertial particles in turbulent boundary layers may be derived from an underlying phase-space probability density function (PDF) equation. These equations, however, are unclosed, and the standard closure approach is to use a quasi-Normal approximation (QNA) in which the fourth moments are approximated as behaving as if the velocities were Normally distributed. Except for particles with weak inertia, the QNA leads to large quantitative errors, and is not consistent with the known asymptotic predictions of Sikovsky (Flow Turbulence Combust, vol. 92, 2014, pp. 41-64) for the moments of the PDF in the viscous sublayer. We derive a new closure approximation based on an asymptotic solution to the transport equations in regions where the effect of particle inertia is significant. The new closure is consistent with the asymptotic predictions of Sikovsky, but applies even outside the viscous sublayer. Comparisons with direct numerical simulations (DNS) show that the new closure gives similar results to the QNA (with the QNA results in slightly better agreement with the DNS) when the viscous Stokes number is $St<10$, but for $St>10$ the new model is in far better agreement with the DNS than the QNA. While the predictions from the new model leave room for improvement, the results suggest that this new closure strategy is a very effective alternative to the traditional QNA approach, and the closure could be refined in future work.

\end{abstract}

\maketitle

\section{Introduction}

The classical model for predicting the concentrations of settling inertial particles in wall-bounded turbulence is that of Rouse \cite{rouse37}. This model is purely phenomenological, and assumes that the effect of particle inertia is negligible except for the finite gravitational settling velocity it introduces. While extensions of this model to particles with small but finite Stokes number have been considered using regular perturbation theory \cite{RichterBLM2018}, developing a model for moderate to large Stokes numbers remains a significant challenge. In order to understand in detail the additional physical mechanisms introduced by finite particle inertia, and therefore the additional terms that an extended Rouse model must capture, in \cite{bragg21}, the settling and concentration profiles of settling inertial particles in wall-bounded turbulence was explored using theory and direct numerical simulations (DNS). The particles were assumed to be small and heavy, with their motion governed by a Stokes drag force and gravity. In contrast to the phenomenological approach of \cite{rouse37}, a rigorous, systematic approach for deriving transport equations for inertial particles in turbulent flows is to derive them as moments of an underlying phase-space PDF equation \cite{reeks91,ZAICHIK1997127,swailes97,reeks_2005,bragg12b}. Therefore, in  \cite{bragg21} the exact (but unclosed) transport equations governing the particle concentration and settling velocities were constructed as moments of a PDF equation for the particle position and velocity. An analysis of the equations led to new insights into the physical mechanisms governing these processes, and how they differ from those in homogeneous turbulence that was explored in \cite{maxey87,tom19}. Data from DNS was then used to evaluate the various terms appearing in these equations, providing insights into the role played by the inertial mechanisms that are absent from the classical model of Rouse \cite{rouse37}. These additional terms were shown in \cite{bragg21} to be of leading order importance in determining the particle settling and concentrations, unless the particle Stokes number is very small, and these terms must therefore be accounted for if a Rouse-type model is to be extended to the case of finite Stokes number particles.

While the analysis of \cite{bragg21} led to new physical insights, in order to develop a predictive theoretical model the hierarchy of moment equations derived from the PDF equation most be closed. The usual closure assumption is to make a quasi-Normal approximation (QNA) \cite{ZAICHIK2010850} (which can also be derived using a Chapman-Enskog approach \cite{SWAILES1998517}), according to which the hierarchy of moment equations is closed by assuming that the fourth moment behaves as if the particle velocities were Normally distributed. Aside from being ad-hoc (in general), this closure approach is known to lead to significant errors in the predictions from the closed moment equations, errors that are both quantitative and qualitative in nature (similar issues also arise when the QNA is used in models of inertial particle-pair transport in isotropic turbulence \cite{bragg14b,bragg14c}). These issues will be discussed in detail in \S\ref{QNA}.

In this paper we explore an alternative closure approximation that captures the asymptotic behavior of the relation between the fourth and second moments of the particle velocity in regimes of the flow where locally the effect of particle inertia is significant. This closure captures the strong non-Gaussianity of the particle velocities in the near-wall region of the flow, and is consistent with the asymptotic behavior of the moments in the viscous sublayer that was described in \cite{sikovsky14,johnson20}. The predictions of the new closed model are compared with DNS data and we find that while the new model is not always in full quantitative agreement with the DNS, it does provide far superior predictions compared to the QNA when the particle inertia is moderate to strong. Moreover, even when there are quantitative discrepancies, the predictions are qualitatively consistent with the DNS data, unlike the QNA model whose solutions also feature a spurious bifurcation near the wall as the Stokes number is increased beyond a threshold value\cite{sikovsky19}. Therefore, while there is still room for improvement, the results suggest that the new closure approach is promising, and could be further refined in future work.

\section{Transport equations for inertial particles in a turbulent boundary layer}
\subsection{Hierarchy of moment equations}
In this work, we consider the transport of small, heavy inertial particles subject to the equation of vertical motion (in what follows, all variables are in wall units, with the usual ``$+$'' superscript omitted for notational simplicity)
\begin{align}
\ddot{z}^p(t)\equiv \dot{w}^p(t)=\frac{1}{St}\Big(u^p(t)-w^p(t)\Big),\label{Veom}      
\end{align}
where $St\equiv \tau_p/\tau_*$ is the particle Stokes number, $\tau_p$ is the particle response time, $\tau_*\equiv \nu u_*^{-2}$ is the fluid time scale based on the friction velocity $u_*$, $z^p(t),w^p(t)$ are the vertical particle position and velocity, and $u^p(t)$ is the vertical fluid velocity at the particle position. The particle volume and mass loadings are assumed sufficiently small to ignore particle-particle collisions and two-way coupling. While this is a highly simplified system, the great difficulties in developing statistical transport equations even for this simple system mean that incorporating additional complexities are best left for future work. Moreover, while as discussed in the introduction, our ultimate interest is in a model for settling inertial particles, we are here focusing on developing an improved closure method for the simpler case of non-settling particles. An extension of the new model proposed in this paper to the case of settling particles will be the subject of a future study.

The joint probability density function (PDF) for $z^p(t),w^p(t)$ in a 2D phase-space with time-independent coordinates $z,w$ is defined as
\begin{align}
\mathcal{P}(z,w,t)\equiv\Big\langle \delta(z^p(t)-z)\delta(w^p(t)-w)\Big\rangle,        
\end{align}
where $\delta(\cdot)$ is the Dirac delta distribution, and the evolution equation is \cite{johnson20,bragg21}
\begin{align}
\partial_t\mathcal{P}=-\nabla_z\Big(\mathcal{P}w\Big)  -\nabla_w\Big(\mathcal{P}\langle\dot{w}^p(t)\rangle_{z,w} \Big),\label{PDFeq}   
\end{align}
where the operator $\langle\cdot \rangle_{z,w}$ denotes an ensemble average conditioned on $z^p(t)=z$, $w^p(t)=w$, and $\nabla_z$ and $\nabla_w$ denote differentiation with respect to $z$ and $w$, respectively.

The $N^{th}$ order moment $\mathcal{M}_N(z,t)$ of the PDF is defined as
\begin{align}
\mathcal{M}_N(z,t)\equiv \int_{\mathbb{R}} w^N\mathcal{P}\,dw=\varrho\mathcal{W}_N,
\end{align}
where $\varrho\equiv \mathcal{M}_0$ is the concentration PDF and $\mathcal{W}_N\equiv\langle[{w}^p(t)]^N\rangle_{z}$ are the moments of the particle vertical velocity. 

The evolution equation for $\mathcal{M}_N(z,t)$ can be obtained from \eqref{PDFeq} and is given by \cite{johnson20}
\begin{align}
\partial_t\mathcal{M}_N=-\nabla_z\mathcal{M}_{N+1} + N\varrho\Big\langle\dot{w}^p(t)[w^p(t)]^{N-1}\Big\rangle_{z}.\label{Meq}
\end{align}
Both terms on the rhs of this equation are unclosed. Closed expressions for the second term can be derived once a closure for $\langle\dot{w}^p(t)\rangle_{z,w}$ is specified. Given \eqref{Veom} we have
\begin{align}
\langle\dot{w}^p(t)\rangle_{z,w} =\frac{1}{St}\Big(\langle u^p(t)\rangle_{z,w}-w\Big),\label{PScurr}      
\end{align}
and various approximations have been introduced for closing the conditional average $\langle u^p(t)\rangle_{z,w}$ (see \cite{bragg12b} for a detailed discussion). The closure approximation that can be proven to be formally consistent with the fully-mixed condition in the limit $St\to0$ is that derived using the Furutsu-Novikov formula assuming that the fluid velocity field has Gaussian statistics \cite{swailes97,bragg12b}, leading to the closure
\begin{align}
\frac{1}{St}\mathcal{P}\langle u^p(t)\rangle_{z,w}\approx \mathcal{P}\kappa -\nabla_z\Big(\mathcal{P}\lambda \Big)-\nabla_w\Big(\mathcal{P}\mu \Big),\label{UavCA}
\end{align}
where $\kappa(z,w,t)$ is a drift coefficient, and $\lambda(z,w,t)$, $\mu(z,w,t)$ are dispersion coefficients for diffusion in $z,w$-space, respectively. The details of these coefficients are not important for the present discussion, and so will be given later, except to note that under standard approximations, their dependence on $w$ is neglected, and $\kappa=\nabla_z\lambda$ is assumed \cite{bragg12} (although this is strictly only valid in the steady-state when $St\to0$ \cite{bragg12b}). These approximations will be assumed throughout this paper, under which \eqref{UavCA} simplifies to
\begin{align}
\frac{1}{St}\mathcal{P}\langle u^p(t)\rangle_{z,w}\approx -\lambda\nabla_z\mathcal{P}-\mu\nabla_w\mathcal{P}.\label{UavCAb}
\end{align}
While the focus of this study is on moment equations derived from a kinetic PDF equation, higher dimensional PDF equations that also include $u^p(t)$ in the phase-space have been considered extensively. For these, $u^p(t)$ is usually modeled via a generalized Langevin model (GLM) \cite{MINIER20011}, which is an approach that was first developed in the context of single-phase turbulent flows \cite{pope}. The kinetic and GLM PDF approaches have their own merits, as discussed in \cite{reeks_2005,reeks18}. Since an ultimate goal of our work is to develop a model that extends that of \cite{rouse37} to the case of finite inertia particles, the kinetic approach is preferred here, since the use of a GLM based PDF equation ultimately requires one to construct the solutions to the moment equations via a Monte-Carlo method, rather than simply as the solution to a set of coupled PDEs (although \cite{VANDIJK20124904} explores direct numerical solutions of the PDEs defined via the GLM PDF model).

\subsection{Quasi-Normal approximation}\label{QNA}

The standard approach for closing the first term on the rhs of \eqref{Meq} is to use a quasi-Normal approximation (QNA) \cite{ZAICHIK2010850} (also derived using a Chapman-Enskog approach \cite{SWAILES1998517}). This approach may be summarized as follows. By specifying the particle acceleration $\dot{w}^p(t)$ that appears in the rhs of \eqref{Meq} using \eqref{Veom}, then the steady-state form of the $N=3$ equation can be re-arranged to give 
\begin{align}
\mathcal{M}_3=-(St/3)\nabla_z\mathcal{M}_4 + \varrho\Big\langle{u}^p(t)[w^p(t)]^{2}\Big\rangle_{z}.\label{M4eq}
\end{align} 
Assuming that $\mathcal{W}_4$ behaves as if $w^p(t)$ were Normally distributed leads to $\mathcal{W}_4\approx 3\mathcal{W}_2^2 $ and hence $\mathcal{M}_4\approx 3\mathcal{M}_2^2/\varrho $. Inserting this into \eqref{M4eq}, and using \eqref{UavCAb} then leads to
\begin{align}
\mathcal{M}_3\approx-St\nabla_z(\mathcal{M}_2^2/\varrho) -St \lambda \nabla_z\mathcal{M}_2.\label{M4eqb}
\end{align}
This equation can then be substituted into the transport equation for $\mathcal{M}_2$, and after some manipulation, this leads to a second-order ODE for $\mathcal{W}_2$ \cite{skartlien07}
\begin{align}
0&=(\mathcal{W}_2+\lambda)\nabla_z^2\mathcal{W}_2+\nabla_z\lambda\nabla_z\mathcal{W}_2 -2\mathcal{W}_2/St^2+2 \mu/St,\label{W2ndorde}
\end{align}
whose solution can be be used to obtain $\varrho$ from the steady-state $N=1$ equation, namely
\begin{align}
0&=-St(\lambda+\mathcal{W}_2)\nabla_z\varrho-St\varrho\nabla_z \mathcal{W}_2.\label{rhoeq}
\end{align}
Since the model assumes that the fluid velocity field is Gaussian when closing $\langle u^p(t)\rangle_{z,w}$, then the QNA is self-consistent in the regime $St\ll1$. However, for $St\geq O(1)$, the statistics of $w^p(t)$ are expected to be strongly non-Gaussian even if the fluid velocity field is Gaussian \cite{sikovsky14}, owing to the non-local nature of the inertial particle dynamics. There are also known to be two particular errors introduced by the QNA, which we now discuss.

First, the QNA leads to behavior for $\mathcal{W}_N$ that is inconsistent with the asymptotic behavior for $z\to 0$ when $St\geq O(1)$ \cite{sikovsky14}. In particular, for $St\to 0$, the scaling of the vertical fluid velocity field for $z\to 0$ implies $\mathcal{W}_N\propto z^{2N}$, and the QNA result $\mathcal{W}_4\approx 3\mathcal{W}_2^2 $ is consistent with this. However, for $St\geq O(1)$, $\mathcal{W}_N\propto z^{\gamma}$ \cite{sikovsky14,johnson20}, where $\gamma(St)$ is the power-law exponent describing $\varrho$ in the limit $z\to 0$, namely $\varrho\sim z^{-\gamma}$. The QNA is not consistent with this because it predicts $\mathcal{W}_4\propto z^{2\gamma}$ rather than the correct behavior $\mathcal{W}_4\propto z^{\gamma}$.

The second issue is that the QNA equation for $\mathcal{W}_2$ given by \eqref{W2ndorde} predicts a bifurcation in the solution as $St$ exceeds a threshold value \cite{sikovsky19}, which through \eqref{rhoeq} also leads to a bifurcation in the solution for $\varrho$. This predicted bifurcation is not supported by DNS data and is argued to be unphysical \cite{sikovsky19}, and will be illustrated in \S\ref{Results}.

In view of these serious issues with the QNA for $St\geq O(1)$, an alternative closure approximation for $\mathcal{M}_4$ is desirable that is both consistent with the known asymptotic behavior of the particle velocities in the limit $z\to0$, and also avoids the unphysical bifurcations predicted by the QNA model.

\subsection{Asymptotic closure approximation}

An alternative closure approximation is motivated by the observation in \cite{johnson20} that the normalized solutions to the steady-state transport equations for $\mathcal{W}_N$ can be written as
\begin{align}
\begin{split}
\mathcal{W}_N\Big/\mathcal{W}_2^{N/2}=\mathcal{C}_N\varrho^{N/2-1}\exp&\Bigg(\frac{N-1}{St}\int^z \mathcal{W}_N^{-1}(q)\Big\langle\Big({u}^p(t)-w^p(t)\Big)[w^p(t)]^{N-2}\Big\rangle_{q}\,dq \\
&\quad- \frac{N}{2St}\int^z\mathcal{W}_2^{-1}\Big\langle u^p(t)\Big\rangle_{q}\,dq\Bigg),\label{Moment_relation}
\end{split}
\end{align}
where $\mathcal{C}_N$ are constants with respect to $z$, but will in general depend on $St$. In view of this result, for $St\gg1$ the quantity $\mathcal{W}_N/\mathcal{W}_2^{N/2}$ behaves asymptotically as
\begin{align}
\mathcal{W}_N\Big/\mathcal{W}_2^{N/2}\sim\mathcal{C}_N\varrho^{N/2-1}\Big(1+O(1/St)\Big).\label{AR}
\end{align}
This asymptotic result is valid for arbitrary $z$, however, it is expected that very large values of $St$ would be needed in practice in order to observe this behavior across the entire boundary layer (in particular, it would require that the Stokes number based on the largest timescale in the flow is $\gg1$). The results in \cite{sikovsky14} also imply that \eqref{AR} is valid in the viscous sublayer even for $St=O(1)$ since for $z\to 0$ the result in \eqref{AR} reduces to the asymptotic results for the regime $St\geq O(1)$ predicted by \cite{sikovsky14}. 

The result in \eqref{AR} yields the asymptotic closure approximation (ACA)
\begin{align}
\mathcal{M}_4\sim\mathcal{C}_4\mathcal{M}_2^2,\label{ARc}
\end{align}
whose most important difference compared to the QNA result $\mathcal{M}_4\approx 3\mathcal{M}_2^2/\varrho$ is the absence of the factor $1/\varrho$. It is precisely because the QNA contains the factor $1/\varrho$ that it leads to a behavior for $\mathcal{W}_4$ that is inconsistent with the asymptotic behavior predicted by \cite{sikovsky14} in the limit $z\to0$. In the near-wall region where $\varrho$ can be very large and exhibits a power-law dependence on $z$ \cite{sikovsky14,johnson20}, the QNA and ACA for $\mathcal{M}_4$ will be radically different, both in terms of their qualitative and quantitative behavior.

In order to use \eqref{ARc} to close the moment equations, the constant $\mathcal{C}_4$ must be specified. While this will in general depend upon $St$, the simplest choice is to use $\mathcal{C}_4=3/\varrho(z_b)$, where $z_b$ is the upper boundary of the solution domain and $\varrho(z_b)$ is the boundary condition imposed when solving \eqref{rhoeq}. In a flow with friction Reynolds number $Re_\tau\to\infty$ and $St$ large but finite, then provided that $z_b$ is large enough to correspond to a height at which the effects of the particle inertia are negligible, $\mathcal{M}_4\sim\mathcal{C}_4\mathcal{M}_2^2$ approaches the QNA result $\mathcal{M}_4\approx 3\mathcal{M}_2^2/\varrho$ as $z\to z_b$. This is a self-consistent choice given that the closure for $\langle u^p(t)\rangle_{z,w}$ assumes that the wall-normal fluid velocities are Normally distributed, and therefore the PDF of $w^p(t)$ should be only weakly perturbed from a Normal distribution in regions where the effect of particle inertia is weak. 

An important point is that although \eqref{ARc} will not be accurate when $St\ll1$, this does not in practice matter. The reason for this is two-fold. First, since
\begin{align}
\mathcal{M}_3=-(St/3)\nabla_z\mathcal{M}_4 + \varrho\Big\langle{u}^p(t)[w^p(t)]^{2}\Big\rangle_{z},\label{M3eq2}
\end{align}
then in the regime $St\ll1$ the contribution from the term involving $\mathcal{M}_4$ (whose closure based on \eqref{ARc} is not accurate for $St\ll1$) will be very small (noting that $\mathcal{C}_4\mathcal{M}_2^2$ and its gradient are finite in the limit $St\to0$). Therefore, errors in the closure for $\mathcal{M}_4$ will only lead to small errors in the overall model predictions for $\mathcal{M}_2$. Second, with the aforementioned choice $\mathcal{C}_4=3/\varrho(z_b)$, then \eqref{ARc} asymptotes to the QNA for small $St$ (for which $\varrho$ is almost uniform), and this is know to yield reasonable predictions for $St\ll1$.

Using \eqref{ARc} to specify $\mathcal{M}_4$ in \eqref{M3eq2}, and substituting the resulting equation for $\mathcal{M}_3$ into the equation for $\mathcal{M}_2$ leads to the second-order ODE for $\mathcal{M}_2$
\begin{align}
0=\mathcal{A}\nabla_z^2\mathcal{M}_2+ \nabla_z\mathcal{A}\nabla_z\mathcal{M}_2-2\mathcal{M}_2/St^2+2\mu\varrho/St,\label{M2eqModel}
\end{align}
where 
\begin{align}
\mathcal{A}\equiv(2\mathcal{C}_4/3)\mathcal{M}_2+\lambda.
\end{align}
Since \eqref{M2eqModel} explicitly contains $\varrho$, then \eqref{M2eqModel} must be solved simultaneously with the equation governing $\varrho$, namely \eqref{rhoeq}. However, we have found that the numerical stability of solutions to the coupled equations for $\varrho$ and $\mathcal{M}_2$ is improved if instead a second-order ODE is solved for $\varrho$. This may be obtained by substituting \eqref{rhoeq} (which comes from the equation for $\mathcal{M}_1$) into the equation for $\mathcal{M}_0$, yielding
\begin{align}
0&=-\lambda\nabla_z^2\varrho-\nabla_z\lambda\nabla_z\varrho-\nabla_z^2 \mathcal{M}_2.\label{rhoeq2}
\end{align}

\subsection{Boundary conditions \& numerical solution}

For the QNA model, two boundary conditions must be specified for $\mathcal{W}_2$. A standard choice is to use $\nabla_z\mathcal{W}_2\vert_{z_a}=0$ and either $\nabla_z\mathcal{W}_2\vert_{z_b}=0$ \cite{sikovsky19} or $\mathcal{W}_2\vert_{z_b}=St\mu(z_b)$, where $z_a,z_b$ are the lower and upper boundary points. The Neumman condition $\nabla_z\mathcal{W}_2\vert_{z_b}=0$ is suitable if $z_b$ lies in the quasi-homogeneous region of the wall-bounded flow, or at the centerline of, e.g. a channel flow. The Dirichlet condition $\mathcal{W}_2\vert_{z_b}=St\mu(z_b)$ is less restrictive since it is appropriate provided that the local equilibrium solution to \eqref{W2ndorde} is accurate, without requiring anything about the gradients of $\mathcal{W}_2$. In the QNA solutions shown later, this Dirichlet boundary condition will be used. The point $z_a$ can be specified as $z_a=d_p/2$, where $d_p$ is the particle diameter.

Since $\varrho$ is decoupled from $\mathcal{W}_2$ in the QNA model, the solution for $\varrho$ can be obtained after obtaining $\mathcal{W}_2$ by solving \eqref{rhoeq}, for which a Dirichlet boundary condition $\varrho(z_b)$ can be used. Given that $\varrho$ is a PDF for $z^p(t)$, its integral over the full flow should be equal to one. Due to the linearity of \eqref{rhoeq}, $\varrho(z_b)$ can be chosen arbitrarily, and the solution can be subsequently re-normalized to satisfy this integral condition. However, if the model is only being solved over a portion of the flow (e.g. the boundary layer), then the absolute values of $\varrho$ cannot be determined, but only the concentration profile relative to some reference value. In this case, the choice of $\varrho(z_b)$ is arbitrary and may be simply set to one.

In the new ACA model, the equations for $\varrho$ and $\mathcal{M}_2$ are coupled, and the boundary conditions should be chosen to be consistent with equation \eqref{rhoeq} which requires 
\begin{align}
\nabla_z \mathcal{M}_2\vert_{z_a}&=-\lambda\nabla_z\varrho\vert_{z_a}.\label{rhoeqBC}
\end{align}
One choice would be to use $\nabla_z \mathcal{M}_2\vert_{z_a}=-\lambda\nabla_z\varrho\vert_{z_a}=0$, and this is the appropriate choice for $St\gg 1$ because $\lim_{St\to\infty}\lambda=0$. For moderate values of $St$, an alternative is to specify $\nabla_z\varrho\vert_{z_a}$ based on the local equilibrium solution to $\varrho$. This is obtained by using the local equilibrium solution $\mathcal{W}_2=St\mu$ in \eqref{rhoeq} yielding $\varrho^{eq}$. With this, the Neumman boundary condition for $\mathcal{M}_2$ is obtained
\begin{align}
\nabla_z \mathcal{M}_2\vert_{z_a}&=-\lambda\nabla_z\varrho^{eq}\vert_{z_a}.\label{rhoeqBC_eq}
\end{align}
This is similar to the approach described in \cite{sikovsky19} to specify $\nabla_z\mathcal{W}_2\vert_{z_a}$ as an alternative boundary condition for the QNA model. However, we found that \eqref{rhoeqBC_eq} can lead to numerical instability of the solution of the ACA model, and therefore we will use $\nabla_z \mathcal{M}_2\vert_{z_a}=0$ for all $St$ values considered. Note that this is consistent with the use of $\nabla_z\mathcal{W}_2\vert_{z_a}=0$ when solving the QNA model.

For the upper boundary, the local equilibrium solution $\mathcal{M}_2\vert_{z_b}=St\mu(z_b)\varrho(z_b)$ may be used. Given the linearity of the equation for $\varrho$, we may use $\varrho(z_b)=1$, and the solution can be subsequently normalized to yield $\int_{z_a}^{z_b}\varrho dz=1$ in the case where the interval $[z_a,z_b]$ spans the height of the whole flow. Since we are using $\nabla_z \mathcal{M}_2\vert_{z_a}=0$, then consistent with \eqref{rhoeqBC} we use  $\nabla_z\varrho\vert_{z_a}=0$ to specify the second boundary condition for \eqref{rhoeq2}.

The QNA and ACA models involve second order, nonlinear ODEs. To solve them, linearization with Newton-Raphson iteration was used. The local equilibrium solution $\mathcal{W}_2(z)=St\mu\vert_{z_b}$ is used as the initial guess (using $\mathcal{W}_2(z)=St\mu(z)$ leads to numerical issues for larger $St$ values), and the solutions converged rapidly, usually within 3 or 4 iterations.

\section{Comparison between models and DNS}\label{Results}
In this section we compare the predictions from the QNA and ACA models for $\varrho$ and $\mathcal{W}_2$ with DNS data of particle transport in an open channel flow. The DNS data is from the same dataset as that in \cite{bragg21}, except that here there is no gravitational settling, and elastic particle-wall collisions are used which leads to a steady state with $\mathcal{M}_1(z)=0 \forall z$. For the transport equations, the dispersion coefficients $\lambda$ and $\mu$ must be specified, and for these we use the standard local approximations \cite{zaichik99,bragg12}
\begin{align}
\lambda(z)&\approx\frac{ \tau_L \langle uu\rangle}{St(1+St/\tau_L)}  ,\\
\mu(z)&\approx \frac{\lambda}{\tau_L},
\end{align}
where $u$ is the vertical fluid velocity at a fixed position (in contrast to $u^p(t)$ which is the vertical fluid velocity along a particle trajectory). In the results that follow, the DNS data for the fluid wall-normal Reynolds stress $\langle uu\rangle$ is used, while the model discussed in \cite{sikovsky19} for the fluid Lagrangian timescale seen by the particle $\tau_L$ was used.

The model equations were solved on a domain $z\in[d_p,z_b]$ with $z_b=200$. Regarding this choice of $z_b$, in the DNS the open channel surface is located at $z=312.5$, and the solutions to the model are insensitive to the choice of $z_b$ for the $St$ values considered if it is chosen in the range $z_b\in(150, 250)$. For $z_b$ significantly outside of this range, the model predictions are compromised because the Dirichlet boundary conditions $\mathcal{W}_2\vert_{z_b}=St\mu(z_b)$  and $\mathcal{M}_2\vert_{z_b}=St\mu(z_b)\varrho(z_b)$ are no longer appropriate, given that they are based on a local equilibrium solution to the equations for $\mathcal{W}_2$ and $\mathcal{M}_2$.

{\vspace{0mm}\begin{figure}
		\centering
		\subfloat[]
		{\begin{overpic}
				[trim = 0mm 60mm 0mm 70mm,scale=0.4,clip,tics=20]{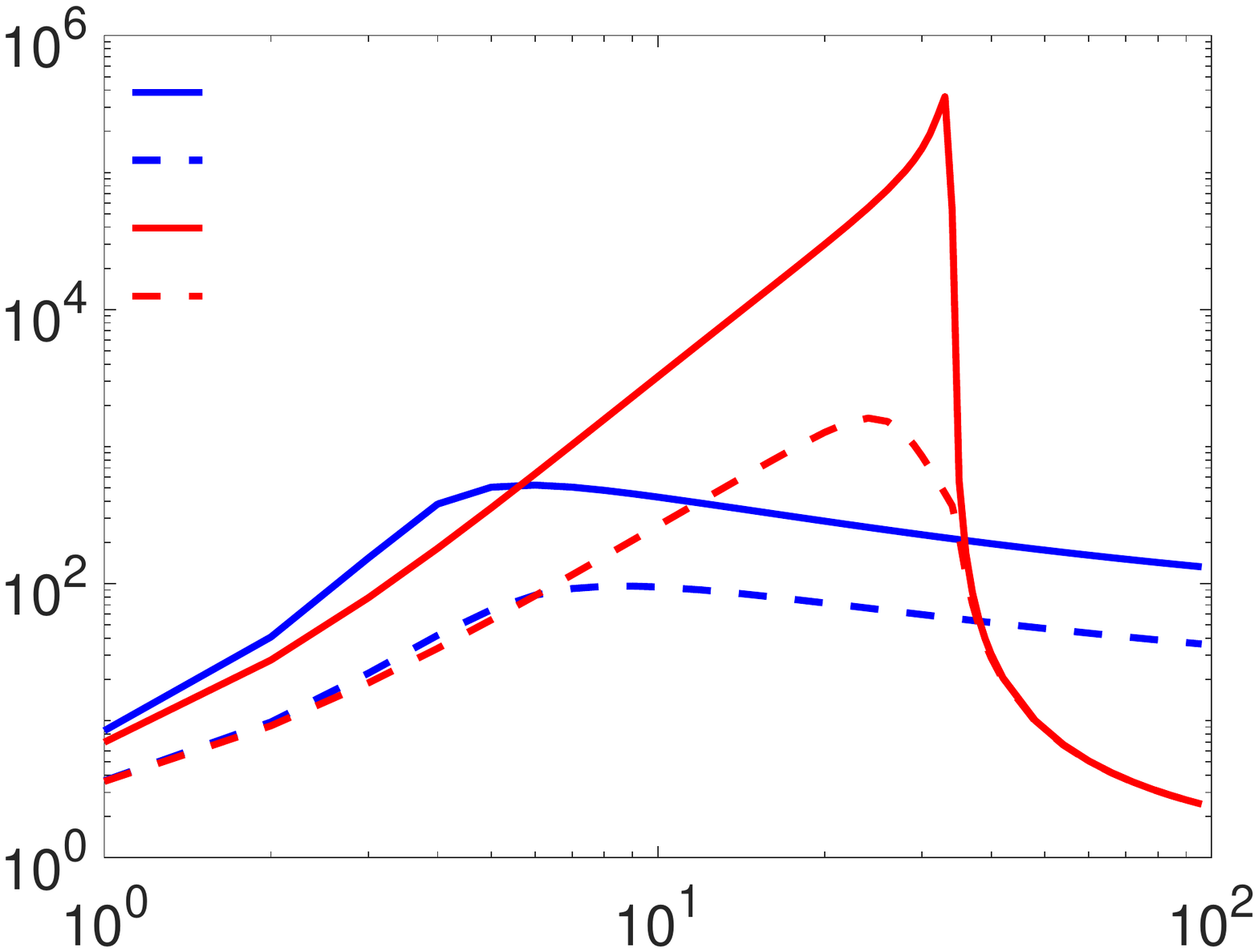}
				\put(120,-3){$St$}
				\put(0,80){\rotatebox{90}{$\varrho(z)$}}
				\put(51,151){ACA, $z=z_a$}	
				\put(51,140){ACA, $z=1$}
				\put(51,129){QNA, $z=z_a$}	
				\put(51,118){QNA, $z=1$}				
		\end{overpic}}
		\subfloat[]
		{\begin{overpic}
				[trim = 0mm 60mm 0mm 70mm,scale=0.4,clip,tics=20]{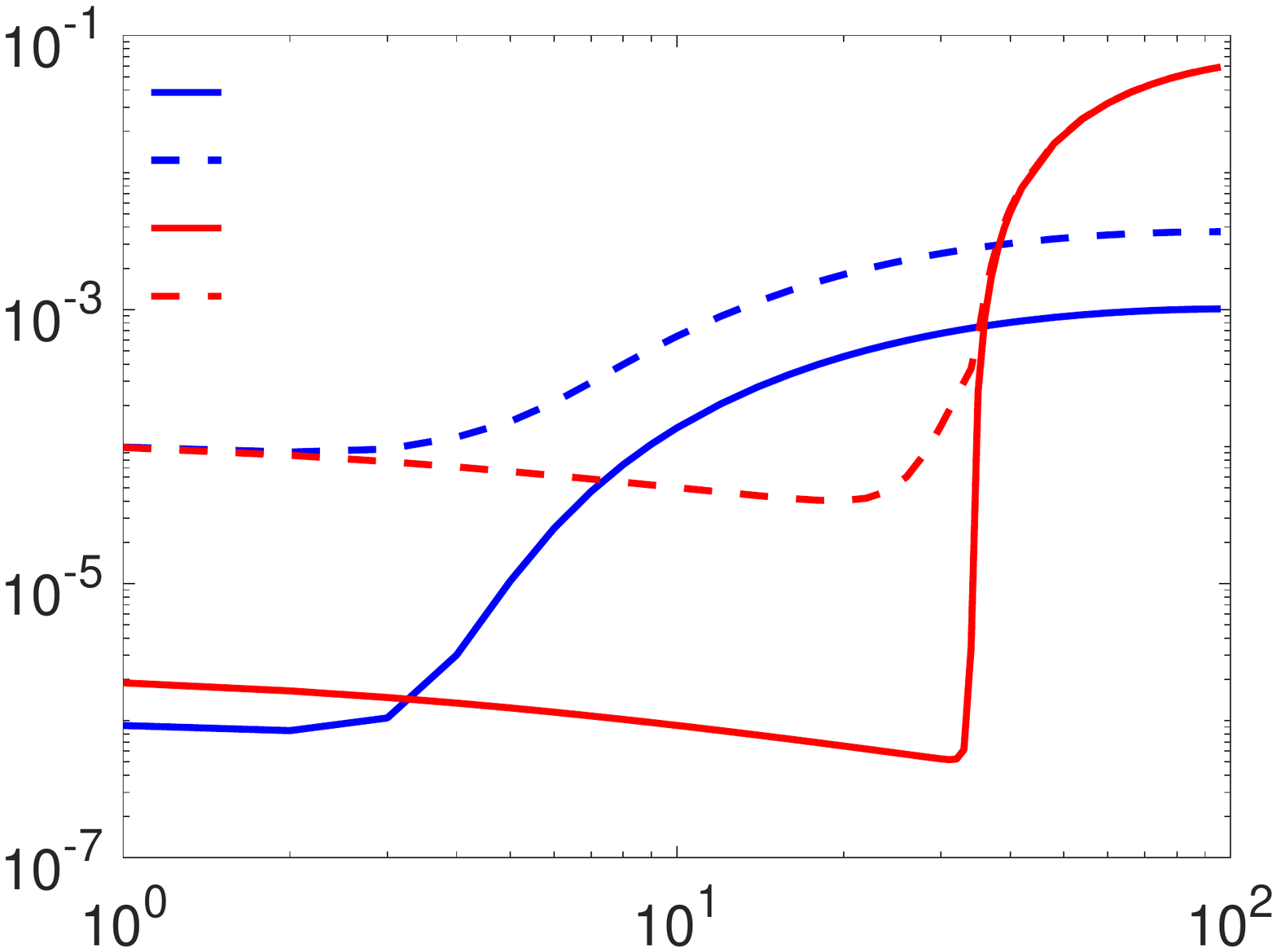}
				\put(120,-3){$St$}		
				\put(-2,77){\rotatebox{90}{$\mathcal{W}_2(z)$}}	
				\put(51,151){ACA, $z=z_a$}	
				\put(51,140){ACA, $z=1$}
				\put(51,129){QNA, $z=z_a$}	
				\put(51,118){QNA, $z=1$}						
		\end{overpic}}
		\caption{Comparison of the predictions from the ACA and QNA models for (a) $\varrho(z)$ and (b) $\mathcal{W}_2(z)$ as a function of $St$ and for $z=z_a$ and $z=1$.} 
		\label{zmin_plots}
\end{figure}}

We begin by comparing the QNA and ACA model predictions with each other, in order to highlight their key differences. In \S\ref{QNA} it was discussed that the QNA model leads to a bifurcation in the solution as $St$ exceeds a threshold value, as also discussed in \cite{sikovsky19}. In figure \ref{zmin_plots} we compare the model predictions for $\varrho$ and $\mathcal{W}_2$ as a function of $St$ and for $z=z_a$ and $z=1$. The results in figure \ref{zmin_plots}(a) show that the QNA model predicts that $\varrho(z_a)$ gradually increases with increasing $St$ until $St\approx 35$, at which point $\varrho(z_a)$ suddenly reduces. The results for $z=1$ show similar behavior, except the drop in $\varrho$ is more gradual. Figure \ref{zmin_plots}(b) shows the associated behavior of $\mathcal{W}_2(z_a)$, for which the QNA model predicts that $\mathcal{W}_2(z_a)$ slowly decreases with increasing $St$ until $St\approx 35$, and then $\mathcal{W}_2(z_a)$ rapidly increases, before slowly increasing with increasing $St$. This is the bifurcation behavior discussed in \cite{sikovsky19}, which appears to be spurious, and is not predicted by the asymptotic analysis of \cite{sikovsky14}. Once this bifurcation occurs, the solutions for $\mathcal{W}_2$ in the viscous sublayer dramatically switch from exhibiting a power-law dependence on $z$ to becoming independent of $z$, which in turn causes a corresponding switch in the behavior of $\varrho$, due to their coupling according to \eqref{rhoeq}. This can be observed in figure \ref{zmin_plots} by noting that for $St> 35$, the solutions for $\varrho$ and $\mathcal{W}_2$ from the QNA model are the same for $z=z_a$ and $z=1$.

By contrast, the predictions from the ACA model do not show such abrupt changes in the behavior of either $\varrho$ or $\mathcal{W}_2$, nor does this model predict that these become independent of $z$. The results do indicate, however, that the ACA model predicts that $\varrho(z)$ as a function of $St$ peaks too early, noting that DNS data suggests the near-wall concentration is strongest somewhere around $St\approx 30$ \cite{MARCHIOLI2008879,johnson20}. This is not surprising, however, given that the ACA is effectively derived for $St\gg1$ as the leading order approximation of an asymptotic series.

{\vspace{0mm}\begin{figure}
		\centering
		\subfloat[]
		{\begin{overpic}
				[trim = 0mm 60mm 0mm 70mm,scale=0.4,clip,tics=20]{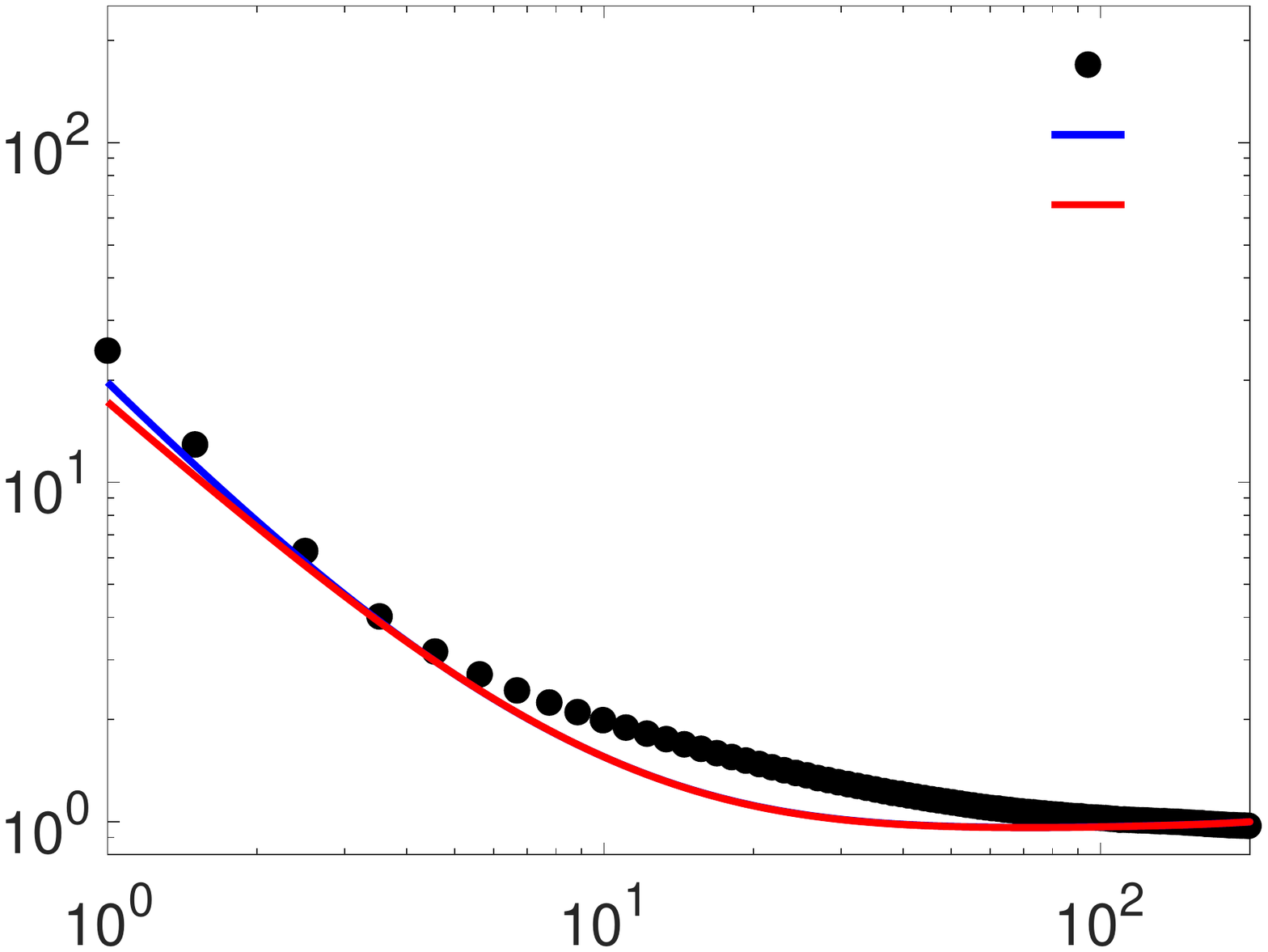}
				\put(130,0){$z$}
				\put(0,70){\rotatebox{90}{$\varrho(z)/\varrho(z_b)$}}
				\put(163,150){DNS}	
				\put(163,138){ACA}	
				\put(163,128){QNA}	
				\put(150,70){$St=2.8$}					
		\end{overpic}}
		\subfloat[]
		{\begin{overpic}
				[trim = 0mm 60mm 0mm 70mm,scale=0.4,clip,tics=20]{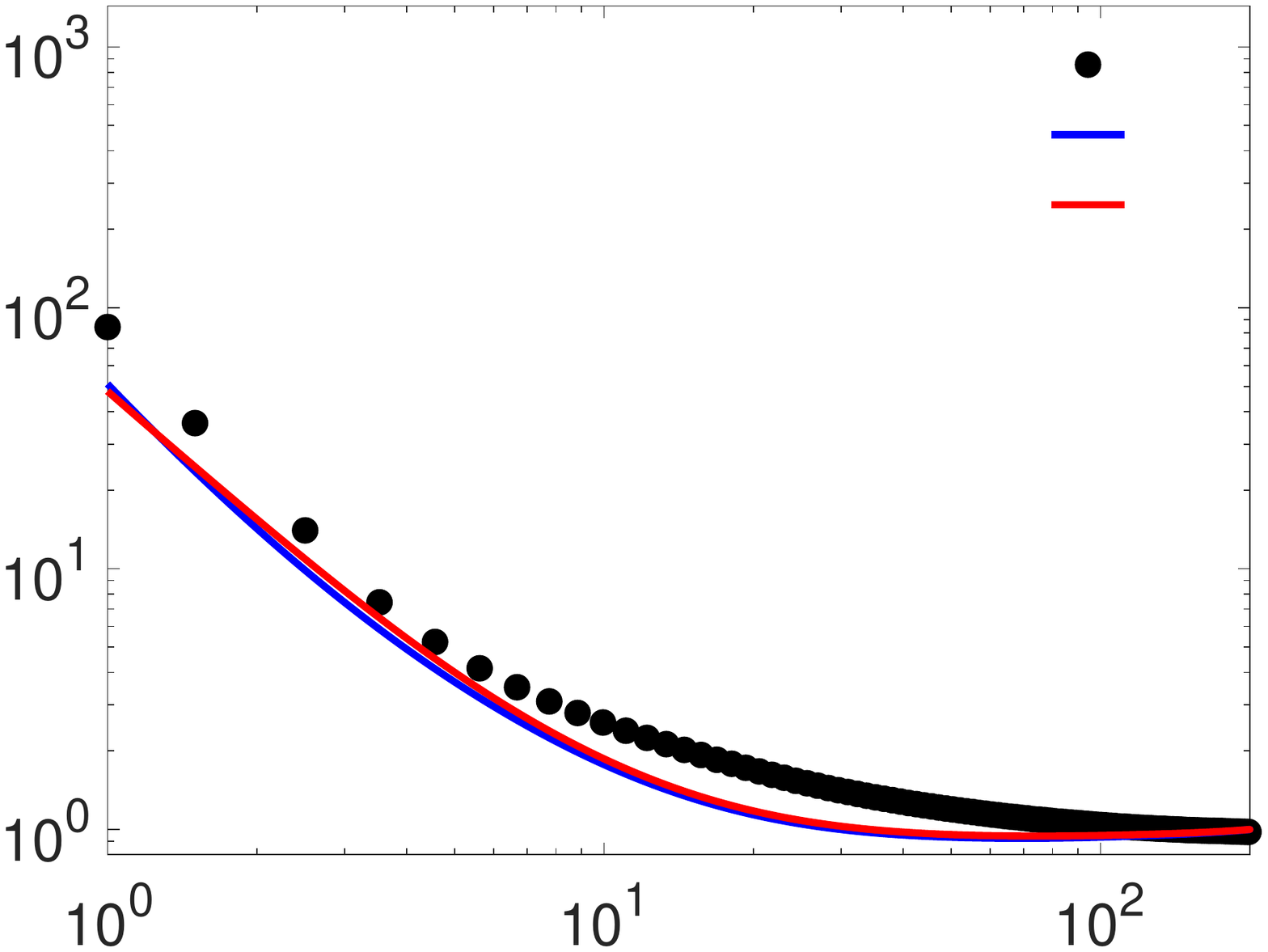}
				\put(130,0){$z$}		
				\put(0,70){\rotatebox{90}{$\varrho(z)/\varrho(z_b)$}}	
				\put(163,150){DNS}	
				\put(163,138){ACA}	
				\put(163,128){QNA}	
				\put(150,70){$St=4.6$}					
		\end{overpic}}
		\\
		\subfloat[]
		{\begin{overpic}
				[trim = 0mm 60mm 0mm 70mm,scale=0.4,clip,tics=20]{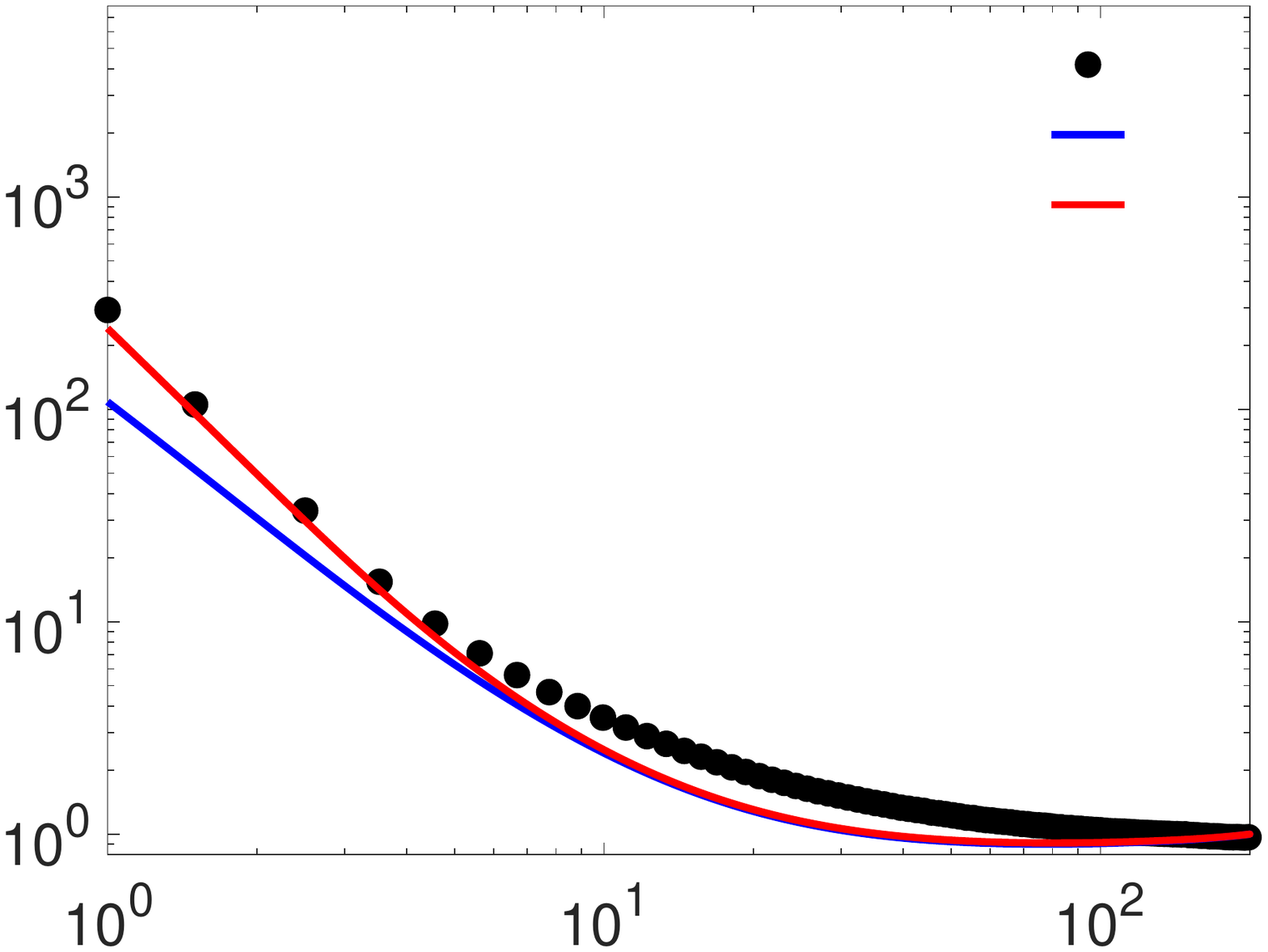}
				\put(130,0){$z$}	
				\put(0,70){\rotatebox{90}{$\varrho(z)/\varrho(z_b)$}}		
				\put(163,150){DNS}	
				\put(163,138){ACA}	
				\put(163,128){QNA}	
				\put(150,70){$St=9.3$}					
		\end{overpic}}
		\subfloat[]
		{\begin{overpic}
				[trim = 0mm 60mm 0mm 70mm,scale=0.4,clip,tics=20]{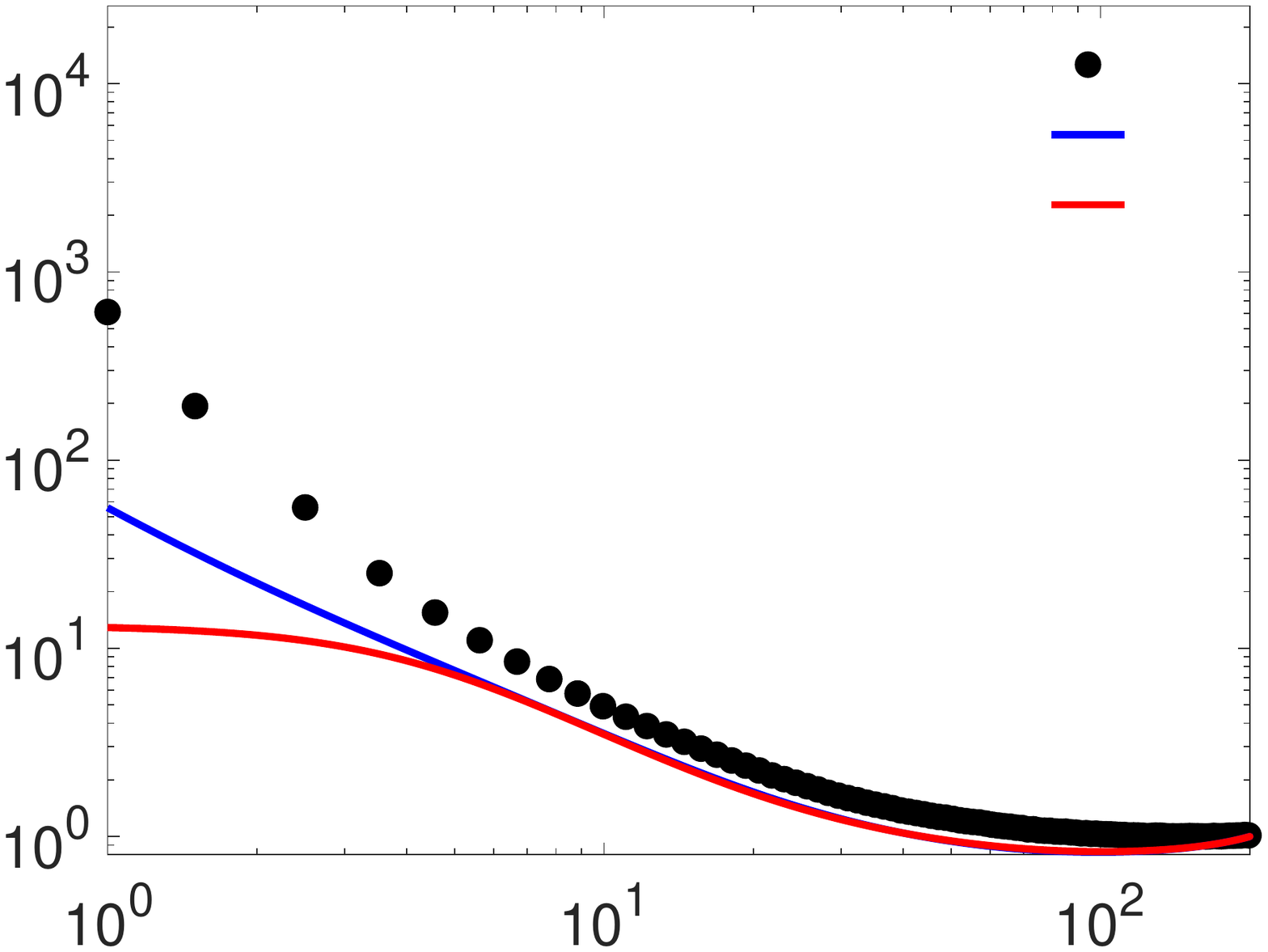}
				\put(130,0){$z$}	
				\put(0,70){\rotatebox{90}{$\varrho(z)/\varrho(z_b)$}}		
				\put(163,150){DNS}	
				\put(163,138){ACA}	
				\put(163,128){QNA}		
				\put(150,70){$St=46.5$}					
		\end{overpic}}	
		\\
		\subfloat[]
		{\begin{overpic}
				[trim = 0mm 60mm 0mm 70mm,scale=0.4,clip,tics=20]{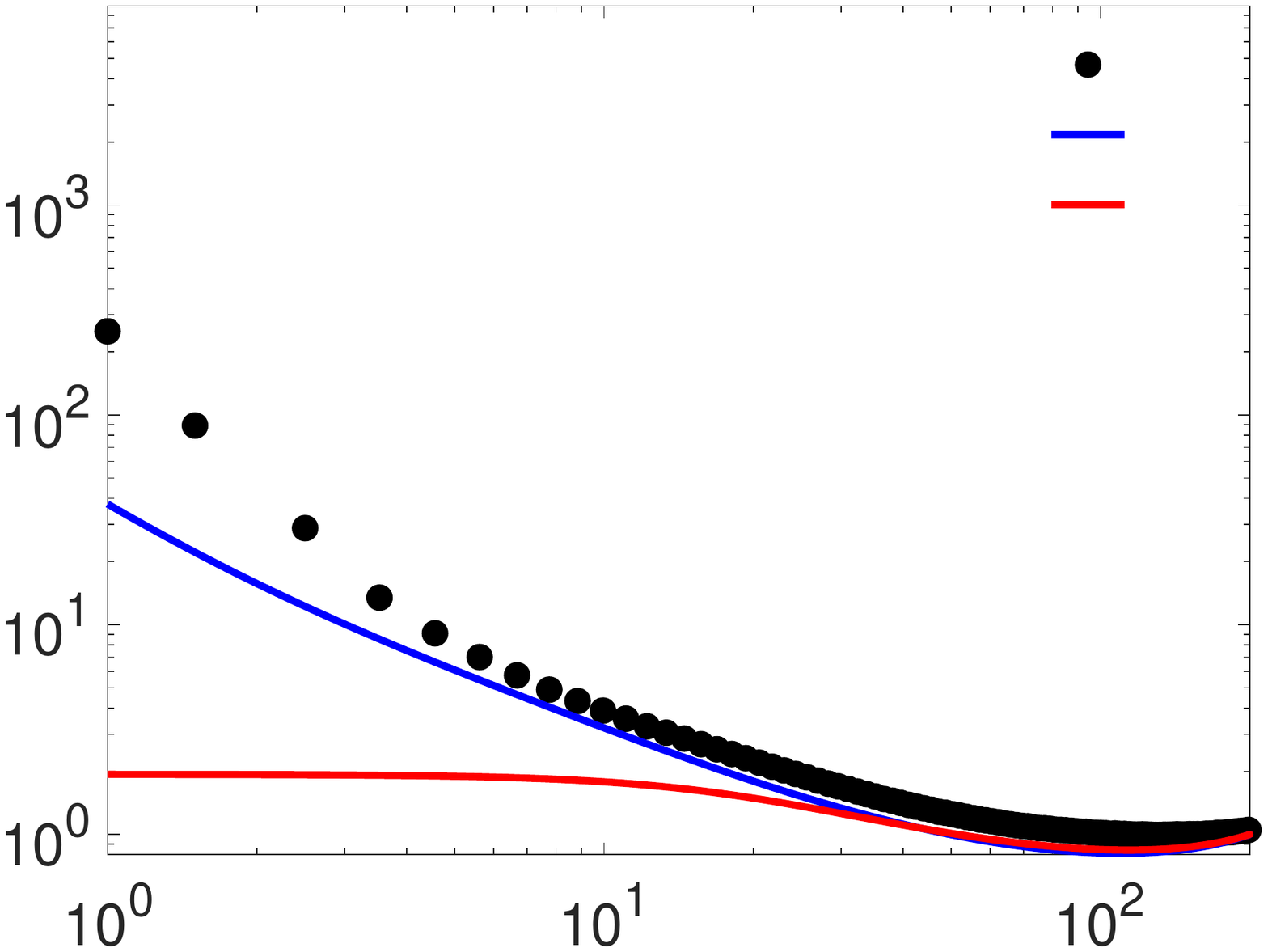}
				\put(130,0){$z$}				
				\put(0,70){\rotatebox{90}{$\varrho(z)/\varrho(z_b)$}}			
				\put(163,150){DNS}	
				\put(163,138){ACA}	
				\put(163,128){QNA}	
				\put(150,70){$St=128$}					
		\end{overpic}}
		\subfloat[]
		{\begin{overpic}
				[trim = 0mm 60mm 0mm 70mm,scale=0.4,clip,tics=20]{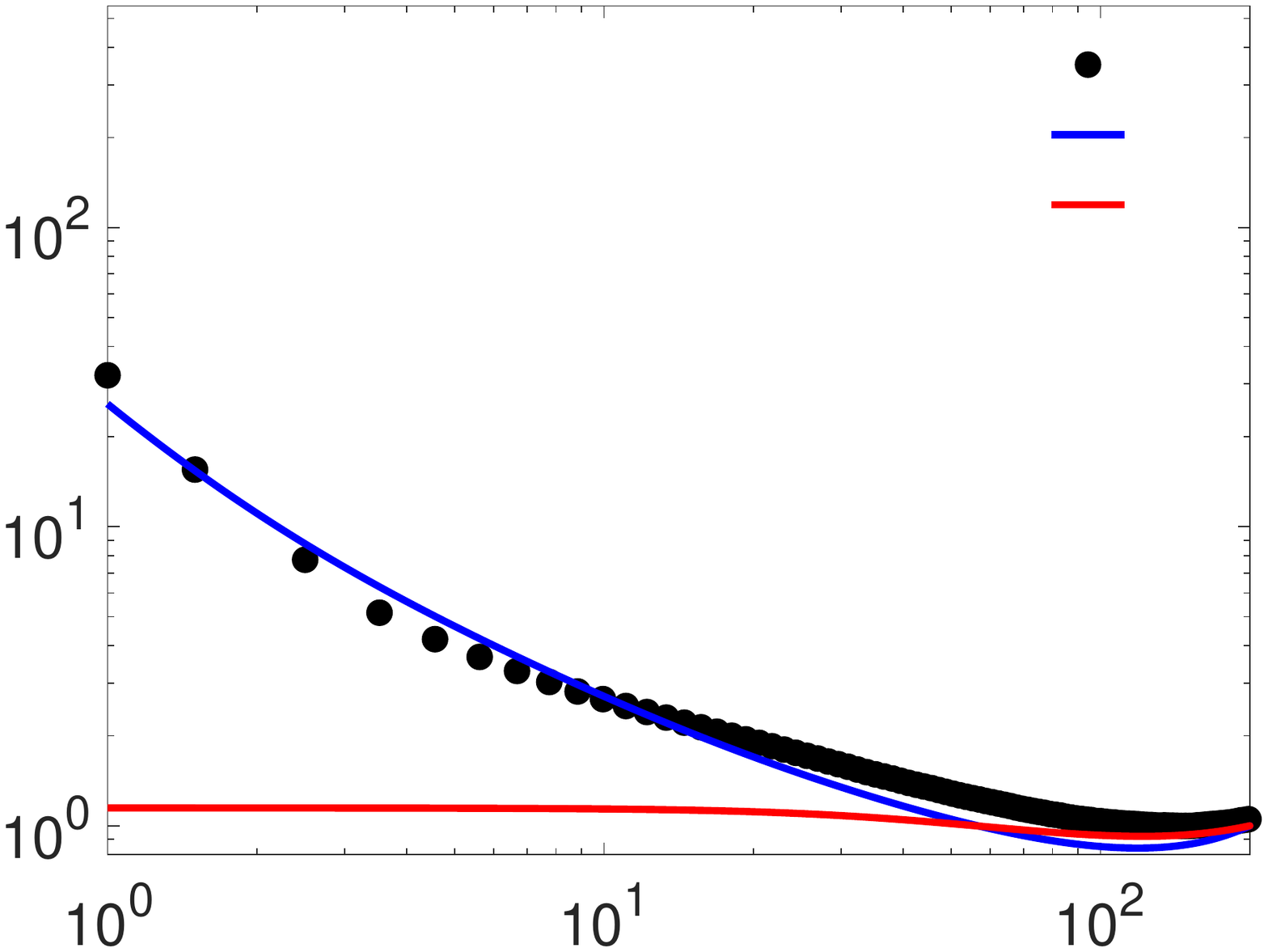}
				\put(130,0){$z$}		
				\put(0,70){\rotatebox{90}{$\varrho(z)/\varrho(z_b)$}}		
				\put(163,150){DNS}	
				\put(163,138){ACA}	
				\put(163,128){QNA}	
				\put(150,70){$St=512$}					
		\end{overpic}}			
		\caption{Comparison of DNS data for $\varrho$ with the predictions from the QNA and ACA for different $St$.} 
		\label{rho_compare}
\end{figure}}

We now compare the model predictions with DNS for $\varrho$ and $\mathcal{W}_2$ and for Stokes numbers $St=2.8, 4.6, 9.3, 46.5, 128, 512$, spanning particles with relatively weak to strong inertia. Figure \ref{rho_compare} shows the results for $\varrho$, and for $St=2.8, 4.6, 9.3$, the QNA and ACA models give similar predictions that are in quite good agreement with the DNS data, with the ACA performing slightly better for $St=2.8, 4.6$, and the QNA performing slightly better for $St=9.3$. Both models slightly underpredict $\varrho$ in the range $7\lesssim z\lesssim 70$. While there are various possible explanations for this, one is that the underpredictions are due to errors introduced by the local approximation for $\lambda$, which can lead to errors for particle transport in turbulent boundary layers \cite{bragg12}. 

Another possibility is that the underpredictions are due to errors in the closure approximation $(1/St)\varrho\langle u^p(t)\rangle_{z}\approx -\lambda\nabla_z\varrho$ that appears in the transport equation governing $\varrho$. In general, the exact expression for $(1/St)\varrho\langle u^p(t)\rangle_{z}$ would involve contributions from infinitely many other terms involving diffusion coefficients and derivatives of $\varrho$ of all orders \cite{bragg21}. That the closure $(1/St)\varrho\langle u^p(t)\rangle_{z}\approx -\lambda\nabla_z\varrho$ only involves a gradient term is a consequence of the assumption that the fluid velocity fluctuations have Gaussian statistics \cite{bragg12}. In \cite{sikovsky14} it was demonstrated that a gradient diffusion closure is asymptotically exact in the viscous sublayer. It might also be expected to be reasonable in and beyond the log-region of a boundary layer where deviations of the fluid velocity statistics from being Gaussian are not expected to be strong. However, in the buffer region where the turbulent production term peaks and where there are intense gradients, the higher-order contributions to $(1/St)\varrho\langle u^p(t)\rangle_{z}$ arising from non-Gaussian fluid velocity fluctuations could be important. This then could explain some of the descrepancies between the model and DNS results for $\varrho$ observed in the region $7\lesssim z\lesssim 70$.

For $St= 46.5, 128, 512$ the QNA has gone past the bifurcation $St$ value discussed earlier, and its predictions for $\varrho$ are in serious error. Not only does it drastically underpredict the values of $\varrho$ in the viscous sublayer, but it also erroneously predicts that $\varrho$ is independent of $z$ in this region. For $St=46.5$, the ACA model underpredicts the DNS data for $\varrho$. 

However, its predictions are closer to the DNS than the QNA, and most importantly, the ACA predicts that for this $St$, $\varrho$ exhibits a power-law dependence on $z$, in agreement with the DNS and asymptotic analysis of \cite{sikovsky14}, but which the QNA fails to reproduce. For $St=128$, the ACA predictions for $\varrho$ are in very good agreement with the DNS down to $z\approx 4$, below which the ACA underpredicts the DNS data. For $St=512$, the ACA is in excellent agreement with the DNS across the range of $z$ considered. The improvement of the ACA predictions as $St$ is increased is of course consistent with the asymptotic nature of its closure approximation.

{\vspace{0mm}\begin{figure}
		\centering
		\subfloat[]
		{\begin{overpic}
				[trim = 0mm 60mm 0mm 70mm,scale=0.4,clip,tics=20]{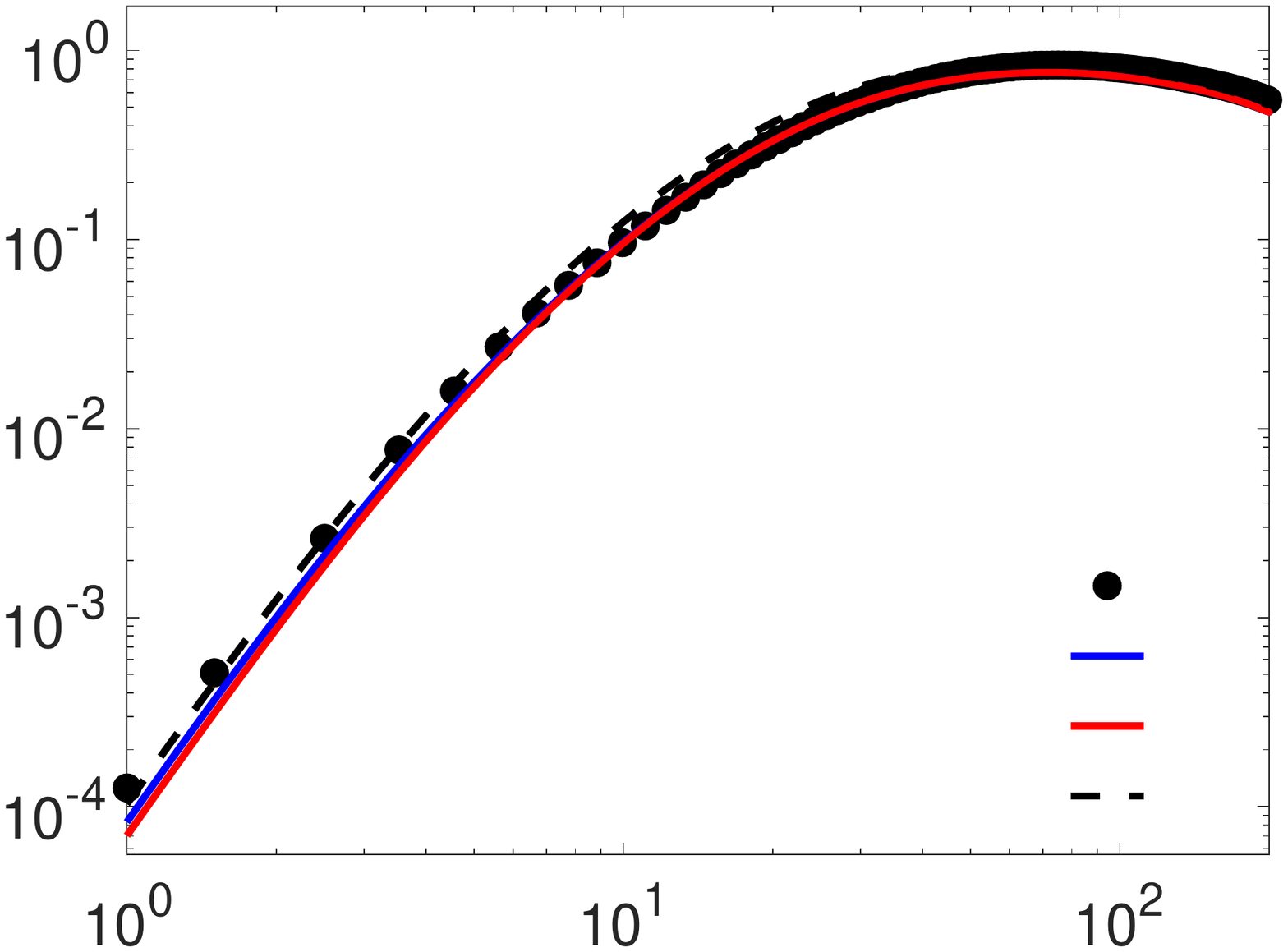}
				\put(130,0){$z$}		
				\put(0,75){\rotatebox{90}{$\mathcal{W}_2(z)$}}		
				\put(163,62){DNS}	
				\put(163,51){ACA}	
				\put(163,40){QNA}		
				\put(163,29){Fluid}					
				\put(150,90){$St=2.8$}				
		\end{overpic}}
		\subfloat[]
		{\begin{overpic}
				[trim = 0mm 60mm 0mm 70mm,scale=0.4,clip,tics=20]{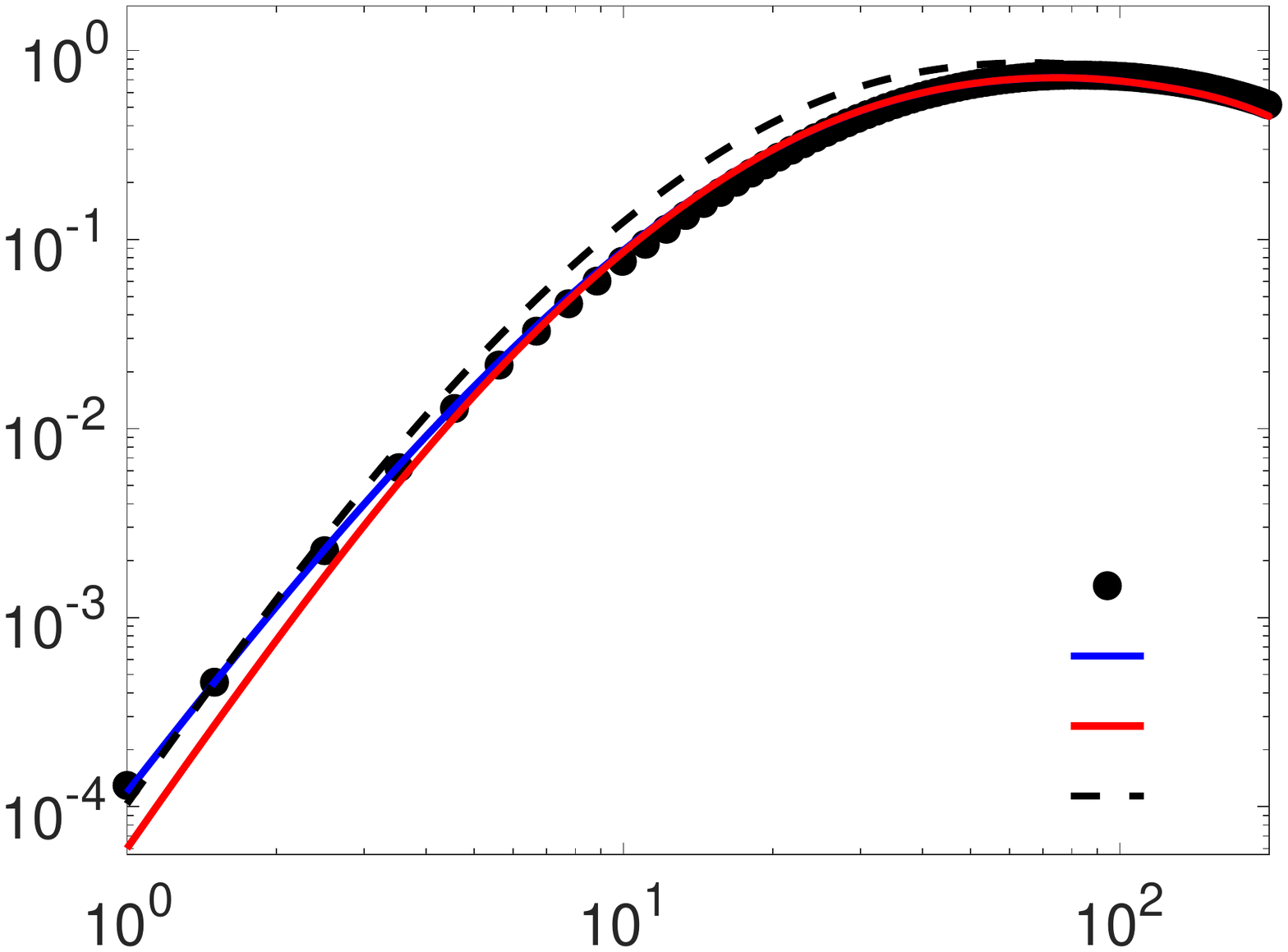}
				\put(130,0){$z$}		
				\put(0,75){\rotatebox{90}{$\mathcal{W}_2(z)$}}		
				\put(163,62){DNS}	
				\put(163,51){ACA}	
				\put(163,40){QNA}		
				\put(163,29){Fluid}	
				\put(150,90){$St=4.6$}					
		\end{overpic}}
		\\
		\subfloat[]
		{\begin{overpic}
				[trim = 0mm 60mm 0mm 70mm,scale=0.4,clip,tics=20]{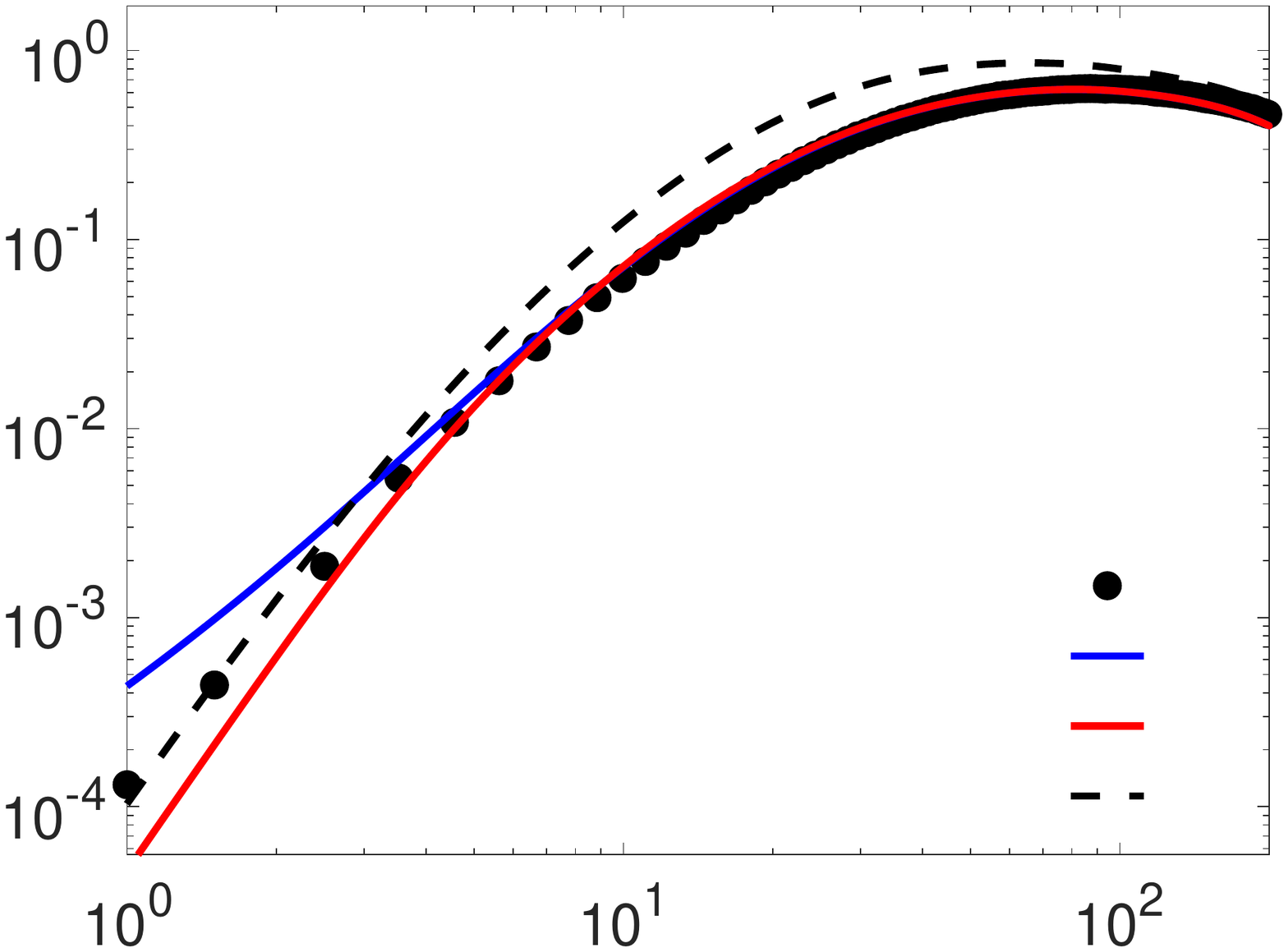}
				\put(130,0){$z$}		
				\put(0,75){\rotatebox{90}{$\mathcal{W}_2(z)$}}		
				\put(163,62){DNS}	
				\put(163,51){ACA}	
				\put(163,40){QNA}		
				\put(163,29){Fluid}	
				\put(150,90){$St=9.3$}					
		\end{overpic}}
		\subfloat[]
		{\begin{overpic}
				[trim = 0mm 60mm 0mm 70mm,scale=0.4,clip,tics=20]{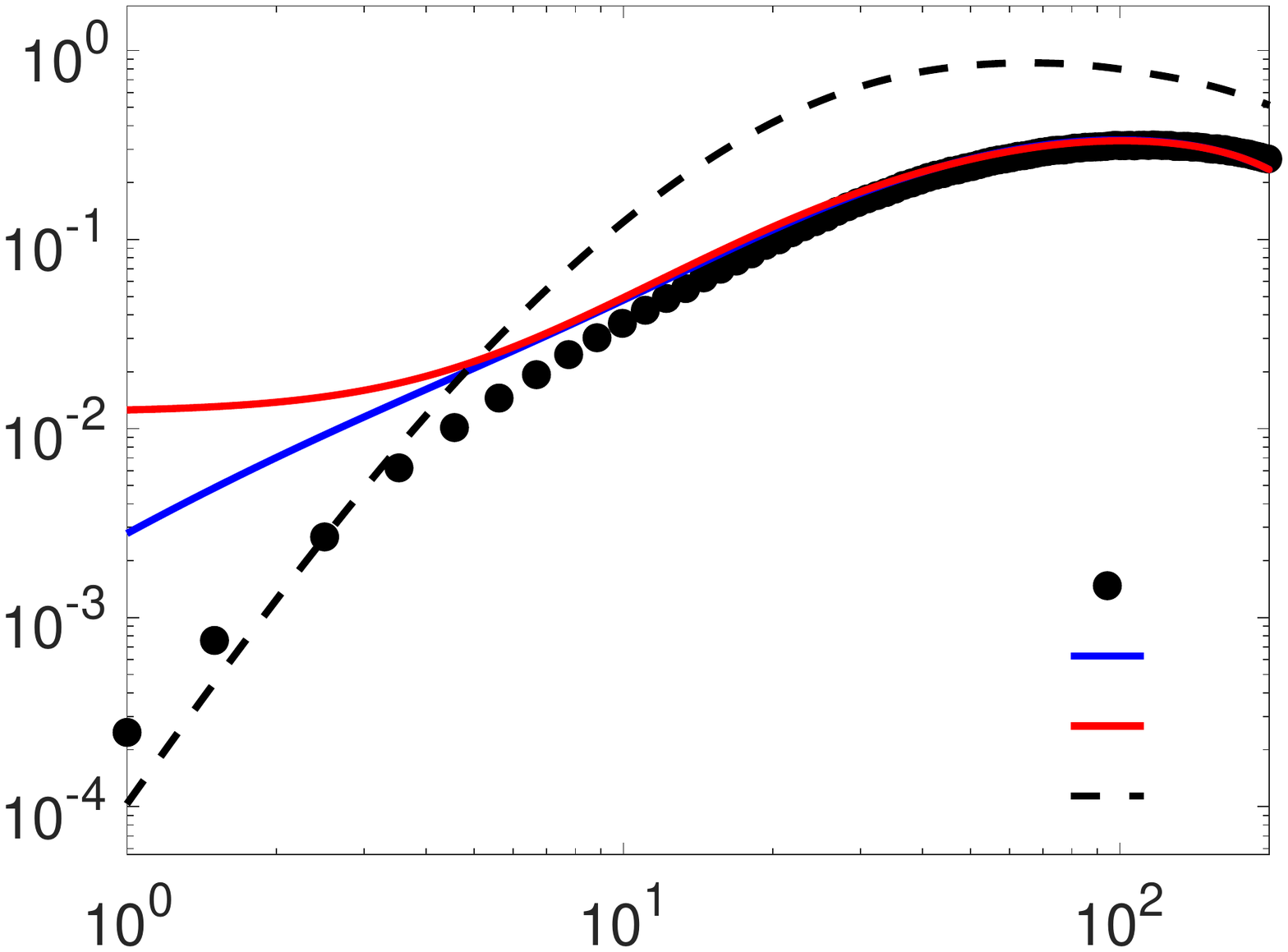}
				\put(130,0){$z$}		
				\put(0,75){\rotatebox{90}{$\mathcal{W}_2(z)$}}		
				\put(163,62){DNS}	
				\put(163,51){ACA}	
				\put(163,40){QNA}		
				\put(163,29){Fluid}		
				\put(150,90){$St=46.5$}					
		\end{overpic}}	
		\\
		\subfloat[]
		{\begin{overpic}
				[trim = 0mm 60mm 0mm 70mm,scale=0.4,clip,tics=20]{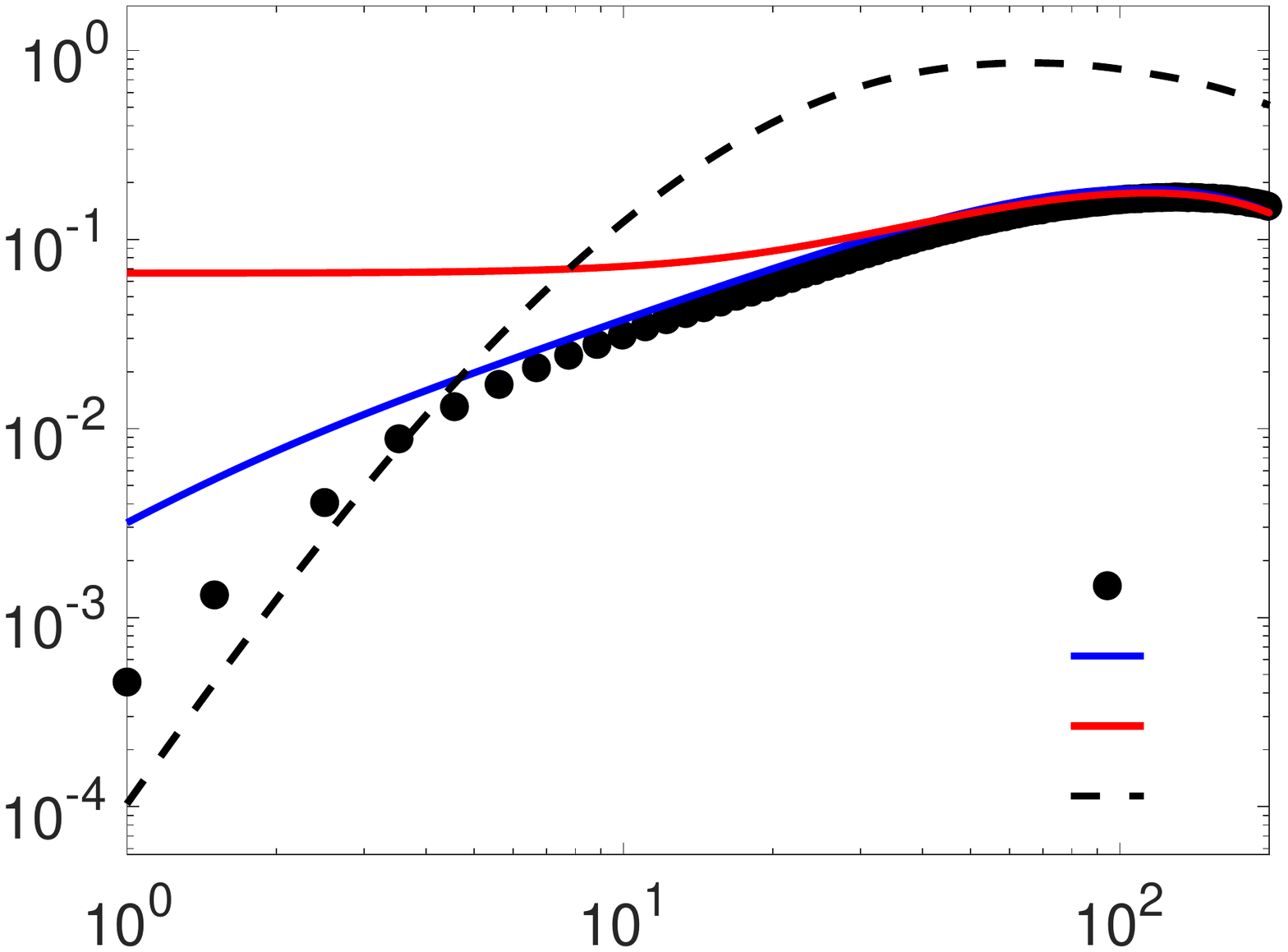}
				\put(130,0){$z$}		
				\put(0,75){\rotatebox{90}{$\mathcal{W}_2(z)$}}		
				\put(163,62){DNS}	
				\put(163,51){ACA}	
				\put(163,40){QNA}		
				\put(163,29){Fluid}	
				\put(150,90){$St=128$}					
		\end{overpic}}
		\subfloat[]
		{\begin{overpic}
				[trim = 0mm 60mm 0mm 70mm,scale=0.4,clip,tics=20]{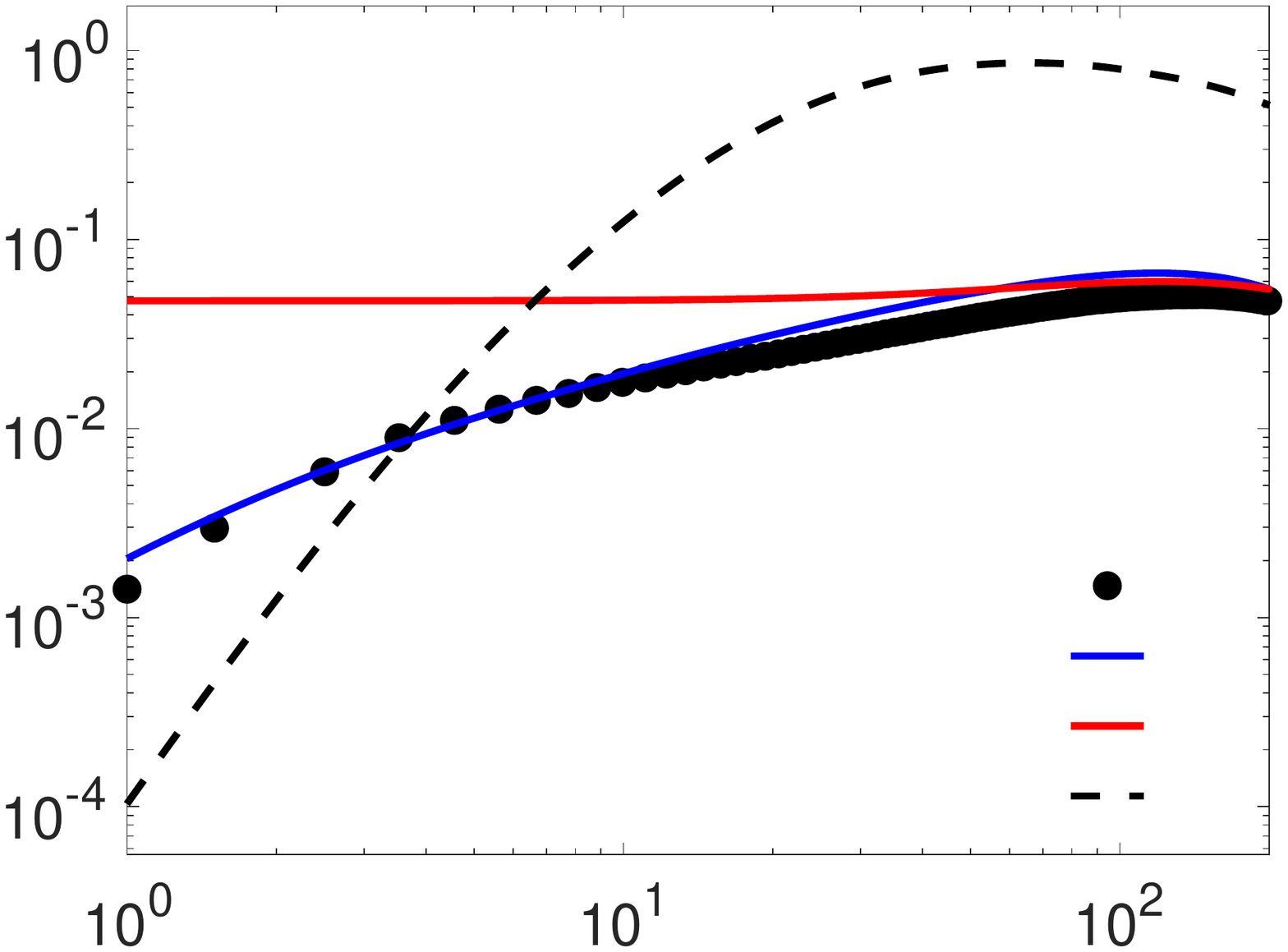}
				\put(130,0){$z$}		
				\put(0,75){\rotatebox{90}{$\mathcal{W}_2(z)$}}		
				\put(163,62){DNS}	
				\put(163,51){ACA}	
				\put(163,40){QNA}		
				\put(163,29){Fluid}	
				\put(150,90){$St=512$}					
		\end{overpic}}			
		\caption{Comparison of DNS data for $\mathcal{W}_2(z)$ with the predictions from the QNA and ACA} 
		\label{W2_compare}
\end{figure}}

\FloatBarrier

Taken together, these results show that the ACA model provides good predictions for $\varrho$ for small and large $St$, but leads to some underpredictions for intermediate $St$. For both intermediate and large $St$, however, the new ACA model provides a significant improvement compared with the traditional QNA, being not only in much better quantitative agreement with the DNS, but also correctly capturing the power-law asymptotic behavior of $\varrho$ in the viscous sublayer, which the QNA does not correctly predict for intermediate and large $St$.

Figure \ref{W2_compare} compares the QNA and ACA model predictions with the DNS data for $\mathcal{W}_2(z)$. The DNS data for the fluid vertical Reynolds stress is also shown for comparison, in order to highlight the extent to which the models capture the effect of the particle inertia on the velocities. For $St=2.8, 4.6$ the QNA and ACA models are both in very good agreement with the DNS, with the ACA model predictions in almost exact agreement with the DNS for $St=4.6$, while the QNA model slightly underpredicts $\mathcal{W}_2(z)$ for $z\lesssim 3$ when $St=4.6$. For $St=9.3$ the QNA and ACA models are both in very good agreement with the DNS down to $z\approx 3$, but below this the QNA model underpredicts the DNS, while the ACA model overpredicts the DNS. For $St=46.5$ the QNA and ACA models are in good agreement with the DNS down to around $z=10$, with both models capturing the strong effects of particle inertia in this regime. However, this $St$ value exceeds the bifurcation $St$ value for the QNA model, and related to this is that the QNA predicts that $\mathcal{W}_2(z)$ becomes constant with values that far exceed those of the DNS at small $z$. The ACA model also significantly overpredicts the DNS at small $z$, but the values are much closer to the DNS than those of the QNA model. Most importantly, while the QNA model predicts that $\mathcal{W}_2(z)$ becomes constant for small $z$, the ACA model preserves the power-law like behavior observed in the DNS. This demonstrates then that despite the quantiative shortcomings of the ACA model at $St=46.5$, it significantly improves upon the QNA model in terms of preserving the right kind of qualitative behavior. For $St=128$, the QNA model is accurate down to around $z=30$, but below this its predictions are in enormous error compared with the DNS, both quantitatively and qualitatively. By contrast, the ACA model is accurate down to around $z=5$. Below this it overpredicts the DNS data, but again preserves a power-like type behavior in this region, which is in much better qualitative agreement with the DNS than the QNA model. Finally, for $St=512$, the QNA model is only accurate down to around $z=100$, and significantly over predicts the DNS data below this, while again predicting that $\mathcal{W}_2(z)$ becomes constant at small $z$, in stark contrast to the DNS. 
{\vspace{0mm}\begin{figure}
		\centering
		\subfloat[]
		{\begin{overpic}
				[trim = 0mm 60mm 0mm 70mm,scale=0.4,clip,tics=20]{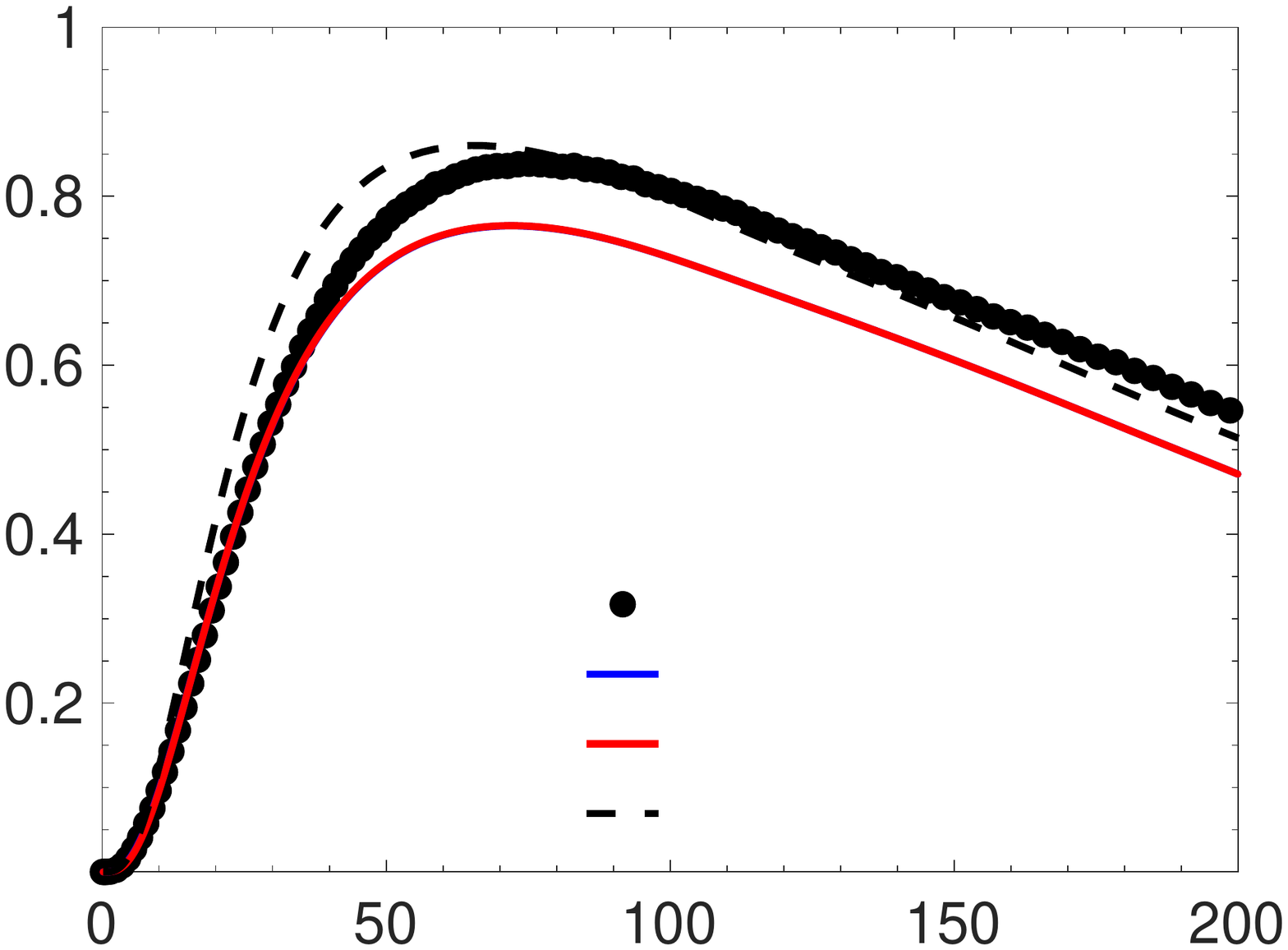}
				\put(130,0){$z$}		
				\put(0,75){\rotatebox{90}{$\mathcal{W}_2(z)$}}		
				\put(130,62){DNS}	
				\put(130,51){ACA}	
				\put(130,40){QNA}		
				\put(130,29){Fluid}			
				\put(160,140){$St=2.8$}					
		\end{overpic}}
		\subfloat[]
		{\begin{overpic}
				[trim = 0mm 60mm 0mm 70mm,scale=0.4,clip,tics=20]{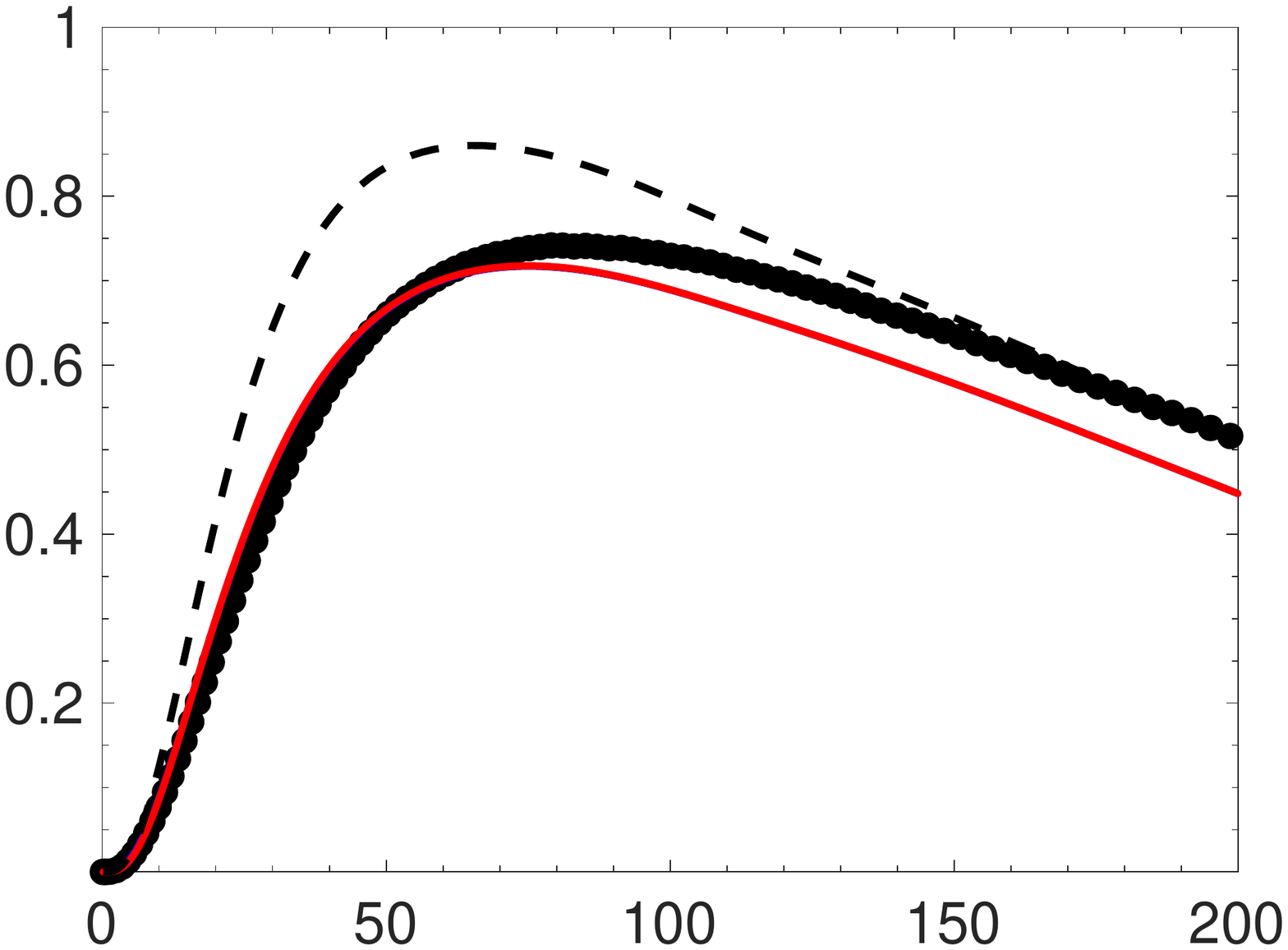}
				\put(130,0){$z$}		
				\put(0,75){\rotatebox{90}{$\mathcal{W}_2(z)$}}	
				\put(160,140){$St=4.6$}					
		\end{overpic}}
		\\
		\subfloat[]
		{\begin{overpic}
				[trim = 0mm 60mm 0mm 70mm,scale=0.4,clip,tics=20]{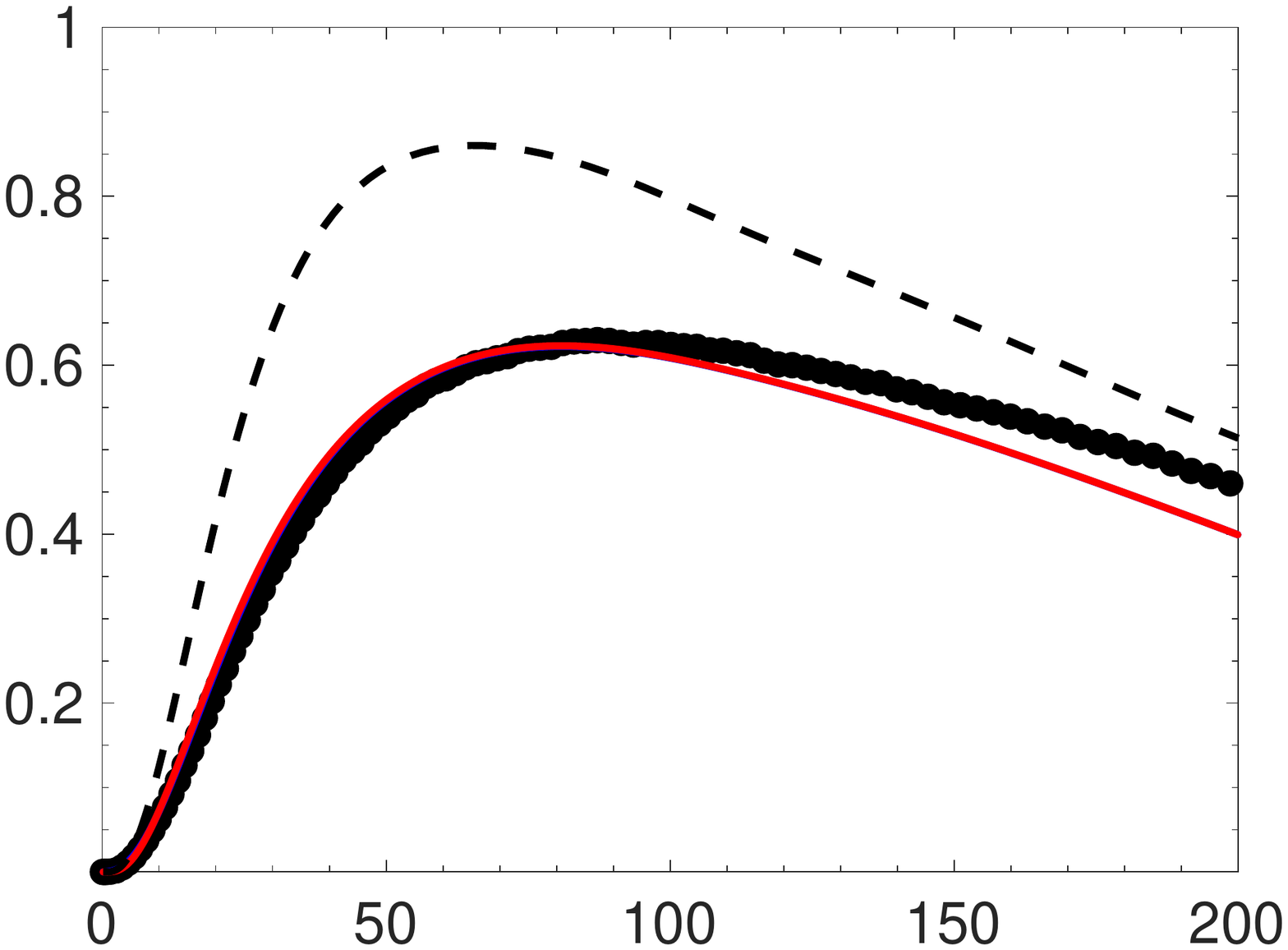}
				\put(130,0){$z$}		
				\put(0,75){\rotatebox{90}{$\mathcal{W}_2(z)$}}		
				\put(160,140){$St=9.3$}					
		\end{overpic}}
		\subfloat[]
		{\begin{overpic}
				[trim = 0mm 60mm 0mm 70mm,scale=0.4,clip,tics=20]{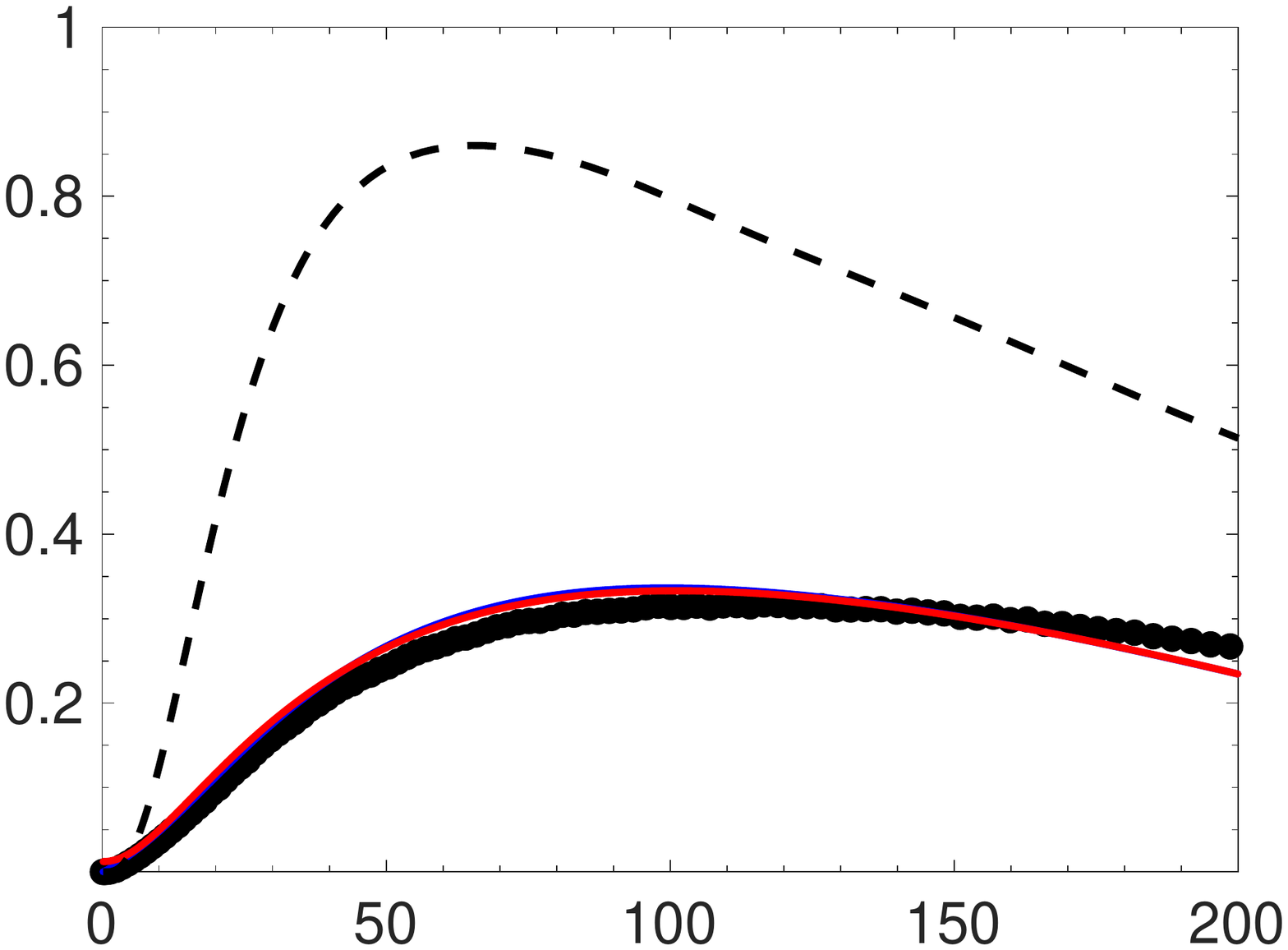}
				\put(130,0){$z$}		
				\put(0,75){\rotatebox{90}{$\mathcal{W}_2(z)$}}	
				\put(160,140){$St=46.5$}					
		\end{overpic}}	
		\\
		\subfloat[]
		{\begin{overpic}
				[trim = 0mm 60mm 0mm 70mm,scale=0.4,clip,tics=20]{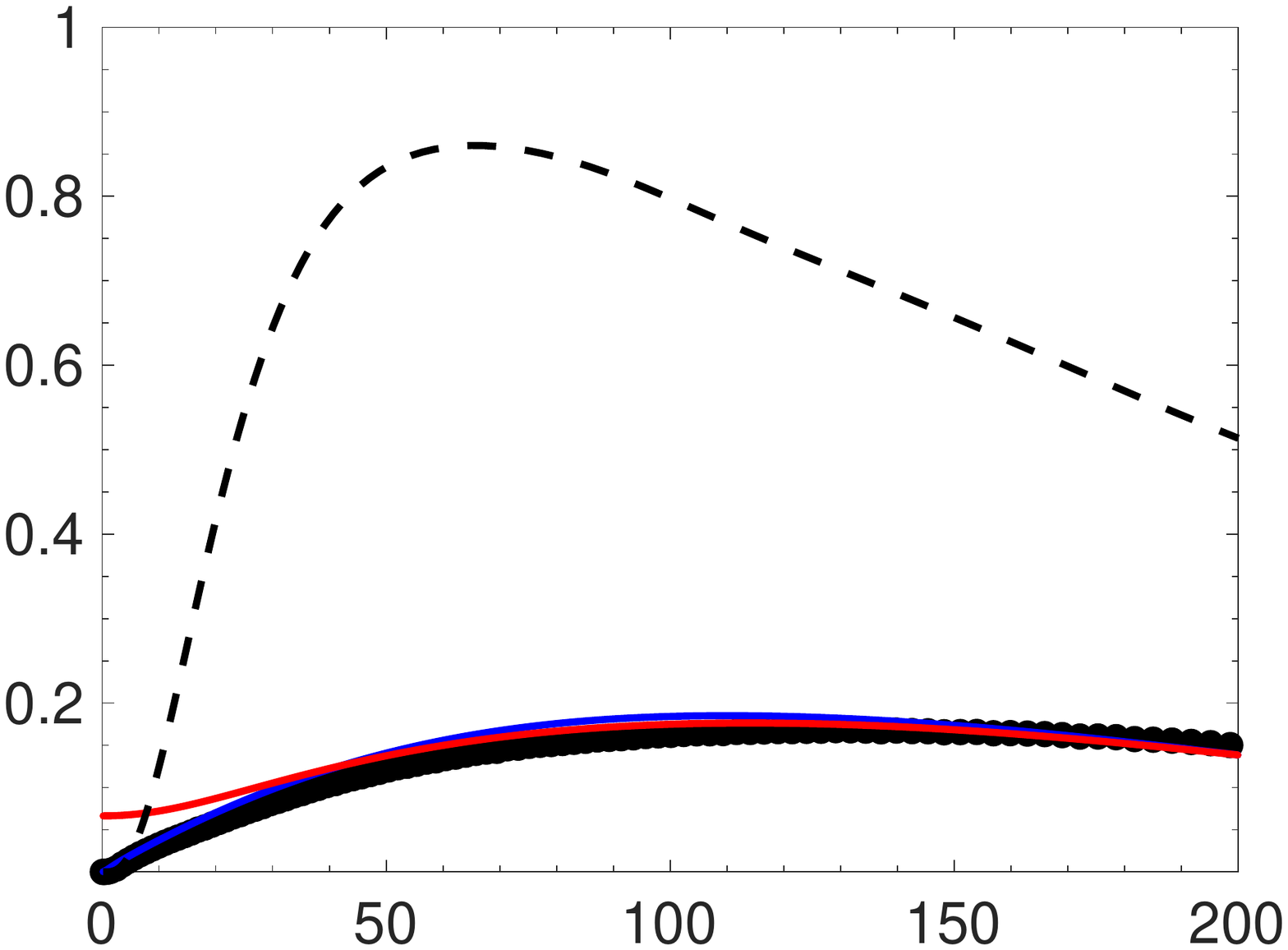}
				\put(130,0){$z$}		
				\put(0,75){\rotatebox{90}{$\mathcal{W}_2(z)$}}	
				\put(160,140){$St=128$}					
		\end{overpic}}
		\subfloat[]
		{\begin{overpic}
				[trim = 0mm 60mm 0mm 70mm,scale=0.4,clip,tics=20]{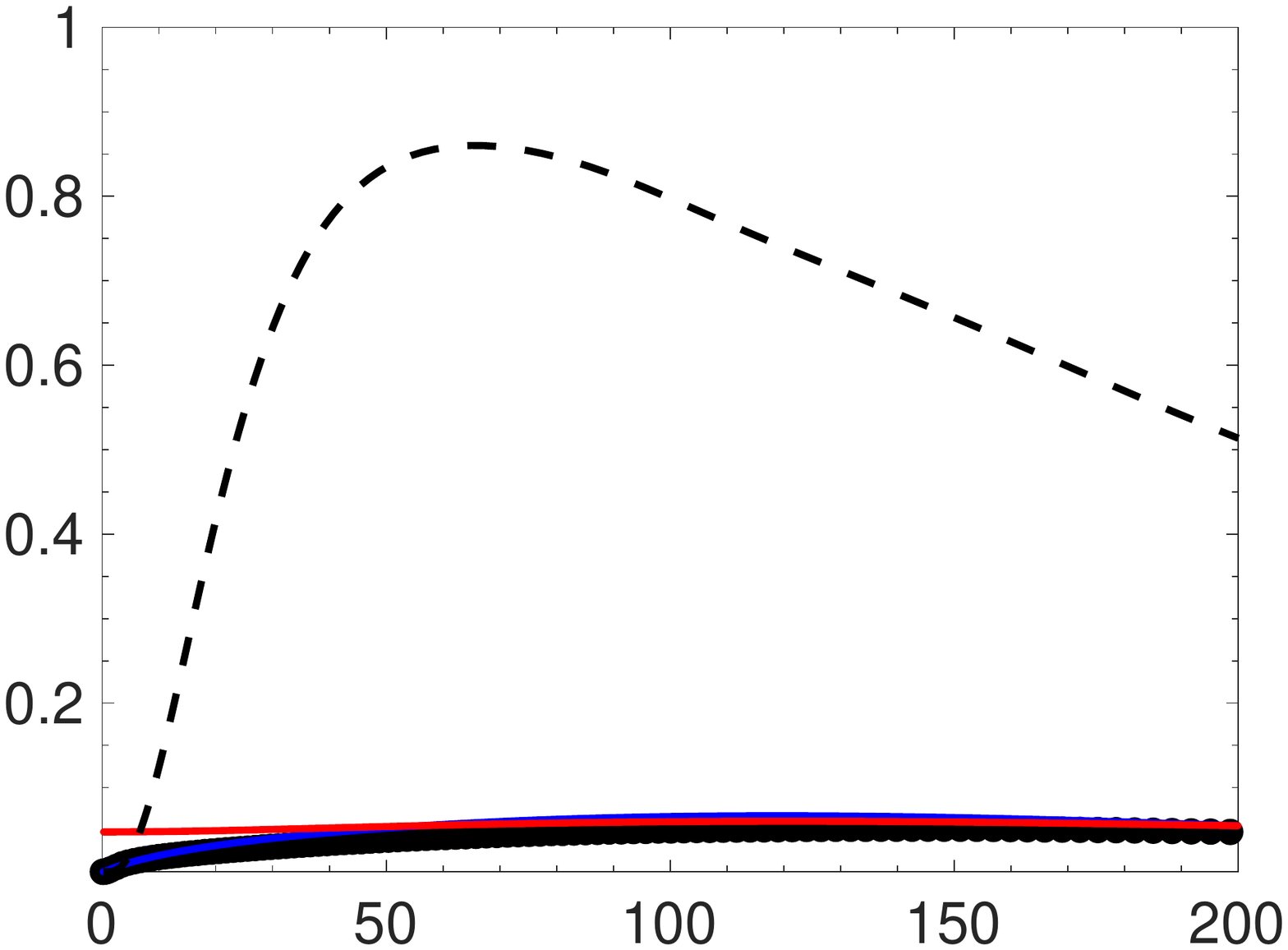}
				\put(130,0){$z$}		
				\put(0,75){\rotatebox{90}{$\mathcal{W}_2(z)$}}		
				\put(160,140){$St=512$}					
		\end{overpic}}			
		\caption{Comparison of DNS data for $\mathcal{W}_2(z)$ with the predictions from the QNA and ACA} 
		\label{Linear_W2_compare}
\end{figure}}
\FloatBarrier
The ACA model slightly overpredicts the DNS for $z>20$, but is in excellent qualitative and quantiative agreement with the DNS below this. Again, the improvement of the ACA predictions compared with the DNS as $St$ is increased is consistent with the asymptotic nature of its closure approximation.

In figure \ref{W2_compare} we again compare the QNA and ACA model predictions with the DNS data for $\mathcal{W}_2(z)$, but this time in a linear scale in order to highlight the behavior at larger values of $z$. At larger $z$, the QNA and ACA predictions are almost identical, and this is because in these regions $\varrho$ does not differ strongly from one, and if $\varrho$ were identically equal to one, then the QNA and ACA closures would be identical. The results show that at greater distances from the wall, e.g. $z\gtrsim 50$, the QNA and ACA models are in general in good agreement with the DNS, with some underpredictions for smaller $St$ that become smaller as $St$ is increased. Comparing the DNS data for $\mathcal{W}_2(z)$ with the fluid Reynolds stress shows that for the range of $St$ considered there is a strong effect of $St$ on $\mathcal{W}_2(z)$, and the models do a very good job of capturing this effect of the particle inertia.

\section{Conclusions}

We have developed a new closure approximation for the moment equations describing inertial particle transport in turbulent boundary layers that are derived from an underlying phase-space PDF equation. Traditionally, a quasi-Normal approximation (QNA) has been used to close the equations, but while this yields good results when the particle Stokes number $St$ is sufficiently small, it leads to significant errors for larger $St$, errors that are both quantitative and qualitative in nature.  We derive a new closure approximation based on an asymptotic solution to the transport equations in regions where the effect of particle inertia is significant. This new closure approximation (referred to as the asymptotoc closure approximation, ACA) differs strongly from the QNA closure in regions where the particle concentration $\varrho$ deviates strongly from being uniform, but asymptotes to the QNA approximation when the concentration is uniform.

Comparisons of the model predictions for $\varrho$ and the variance of the vertical particle velocity $\mathcal{W}_2$ with DNS data show that while the QNA and ACA model make similar predictions at smaller $St$ that are in good agreement with the DNS, their predictions differ dramatically at larger $St$. The ACA model predictions are in good agreement with the DNS over a much wider range of the boundary layer. At smaller distances from the wall, even when the ACA model predictions are not in quantitative agreement with the DNS, they correctly preserve the power-law like behavior of $\varrho$ and $\mathcal{W}_2$, unlike the QNA model that erroneously predicts that these functions become independent of $z$. For very large $St$, the ACA model is in excellent quantitative agreement with the DNS data. The new ACA model therefore dramatically improves on the traditional QNA model.

In order to address the remaining quantitative deficiencies of the ACA model, two possibilities should be explore in future work. First, the coefficient $\mathcal{C}_4$ that appears in equation \eqref{ARc} was obtained by enforcing that in regions of the boundary layer where the particle inertia is weak that the ACA closure asymptotes to the QNA closure. This yields a value for $\mathcal{C}_4$ that is independent of $St$, whereas in reality it probably should depend on $St$. Improving the specification of $\mathcal{C}_4$ to include an appropriate $St$ dependence could improve the accuracy of the ACA model. Second, the ACA closure is formally obtained as the leading order term in an asymptotic series for the regime $St\gg1$. It may be possible to improve upon this by either incorporating the next term in the expansion, or perhaps by using a renormalization approach to perform a partial summation of some of the terms in the series. This will be explored in future work, together with extensions of the model to include the effect of gravitational settling.

\section{Acknowledgements}

Support from the Army Research Office (award $\#$ W911NF-22-2-0222) is gratefully acknowledged. The computing resources and services used in this work were provided by the VSC (Flemish Supercomputer Center), funded by the Research Foundation - Flanders (FWO) and the Flemish Government.

\bibliography{bib/references}

\end{document}